\documentstyle[12pt,aaspp]{article}
\eqsecnum
\begin{document}

\title{Nonlinear Instabilities in Shock-Bounded Slabs}

\author{Ethan T. Vishniac}

\affil{I: ethan@astro.as.utexas.edu, Department of Astronomy, University of
Texas, Austin, TX 78712}

\begin{abstract}
We present an analysis of the hydrodynamic stability of a cold
slab bounded by two accretion shocks.
Previous numerical work (\cite {hun86,ste92}) has shown that when the
Mach number of the shock is large the slab is unstable.
Here we show that to linear order both the bending and breathing
modes of such a slab are stable, with a real frequency of
$c_sk$, where $k$ is the transverse wavenumber. However, nonlinear
effects will tend to soften the restoring forces for bending
modes, and when the slab displacement is comparable to its
thickness this gives rise to a nonlinear instability. The growth
rate of the instability, above this threshold but for small
bending angles, is
$\sim c_sk (k\eta)^{1/2}$, where $\eta$ is the slab displacement.
When the bending angle is large (i.e. $k\eta$ of order unity)
the slab will contain a local vorticity
comparable to $c_s/L$, where $L$ is the slab thickness.
We discuss the relationship between this work
and previous studies of shock instabilities, including the implications
of this work for gravitational instabilities of slabs.  Finally,
we examine the cases of a decelerating slab bounded by a single shock
and a stationary slab bounded on one side by thermal pressure.  The
latter case is stable, but appears to be a special case.  The former
case is subject to a nonlinear overstability driven by deceleration
effects.  We conclude that shock bounded slabs
with a high density compression ratio generically produce substructure
with a strong local shear, a bulk velocity dispersion like the sound
speed in the cold layer and a characteristic scale comparable to the
slab thickness.  We discuss the implications of this work for cosmology
and the interstellar medium.
\end{abstract}


\section{Introduction}

Strong shock waves are an important dynamical element of the
interstellar medium.  They affect the distribution and thermal
state of gas in the galactic disk and may have an important
influence on the rate of star formation in dense molecular
clouds (for a general review see \cite {dm93}).  Moreover,
in the hot dark matter (HDM) model of galaxy formation the dynamics
of shocks are an important part of the process of galaxy formation
(\cite {y91}).  For many of
these processes, as well as for shock diagnostics, it is
important to consider the stability of the shock fronts.  For
a purely hydrodynamic shock considered as a surface dividing
two semi-infinite spaces the stability of the shock can
be proven under nearly all circumstances (\cite{e62}).
However, realistic shocks in the interstellar medium, or in
the early universe, are expected to have a complicated postshock
structure.  This structure will be affected by the details of
radiative cooling and heating, by the presence (at least in the
ISM) of magnetic fields, and by the history and dynamics of
the shock itself (i.e. the finite thickness of the postshock
gas and its overall deceleration or acceleration).  Each
of these elements may lead to instabilities (cf. section 5 in
Draine and McKee (1993) and references therein).

In this paper we will consider the stability of a purely hydrodynamic
(i.e. collisionally dominated and unmagnetized) slab confined
by a shock on at least one surface.  Numerical simulations of
slabs confined between shocks of equal strength (\cite{hun86,ste92})
indicate the presence of a global hydrodynamic instability of an
unknown nature.  We will show that this instability is a nonlinear
effect which can arise whenever the shock compression ratio
is large.  Hunter et al. (1986) referred to this instability as a
Rayleigh-Taylor Instability, which is an unrelated linear effect.
Stevens et al. (1992) referred to it as an example of a ``thin shell
instability''.
Here we refer to it as the ``Nonlinear Thin Shell Instability''
(NTSI) in order to follow their notation while still distinguishing it
from the purely linear effect first described in Vishniac (1983).

We will not consider the effects of magnetic or cooling instabilities here.
Under appropriate circumstances these can be expected to provide
the raw material for the purely hydrodynamic instability we will
explore in this paper (and perhaps vice-versa).  However, the
interplay between these processes is sufficiently complicated that a
numerical approach is probably required.

In section 2 we will discuss the case of a case of a slab bounded
on either side by accretion shocks of equal strength.  We present
a formalism designed to capture the major features of the flows
within the slab, and show that a strong nonlinear instability is
expected to set in when the slab position is slightly perturbed.
We also discuss the limits of this formalism and give a
dimensional analysis of our results.  In section 3 we extend
this to a stationary slab bounded on one side by a shock
and a decelerating slab bounded on one side by a shock.
In section 4 we discuss the linear theory of the gravitational
instabilities of thin slabs and the implications of the current
work for star formation and galaxy formation within the HDM
model.  Finally, in section 5 we summarize our results and
compare them to earlier work on the hydrodynamic instabilities
of strong radiative shocks.

\section{The Shock Bounded Slab}

We are concerned with the stability of an inviscid fluid slab confined
between two accretion shocks.  For simplicity we will begin by
assuming that the unperturbed slab is motionless and contains
material accreted from the winds that confine it.
We will also assume that the fluid is isothermal. Since we are
concerned with the possibility of a global instability we
will aim at the simplest possible description of the flows within
the slab. In practice this means that we will consider fluid
quantities that are averaged through the slab while also
allowing for some large scale shear within the slab.  Flows on
smaller spatial scales will be included only as a source of
turbulent dissipation.  We will consider the kinds of errors
this may introduce later on in this section.

To start with we will define our unperturbed state as a motionless
slab with a sound speed $c_s$ confined between $z_1$ and $z_2$.
We take $\hat z$ as the coordinate direction perpendicular to the unperturbed
slab and $\hat x$ as the transverse coordinate.
The external gas moves towards the slab from both directions
with a velocity $\pm V_E\hat z$ and a
density $\rho_E$.  This gives a shock speed of $V_s$ such that
$V_E=V_s-c_s^2/V_s$ and a slab density of $\rho_s=\rho_E V_s^2/c_s^2$.
The slab has a total thickness of $L=2c_s^2t/V_s$.

It will be useful to start by considering the important physical effects
we want to model.  Suppose we consider a perturbation where the midplane
of the slab is displaced as shown in Fig. 1 (i.e. a ``bending mode'').
Gas striking the slab at point A will move upwards and to the left.
Gas striking the slab at point B will downwards and to the right.
No net transverse mass flow will result, but there will be a net
transport of $\hat z$ momentum in the $\hat x$ direction.  This
effect is nonlinear, but being quadratic in velocity, has the
same sign on opposing faces of the slab.  These terms will tend to
deposit positive $\hat z$ momentum in those parts of the slab which
are displaced upwards, whereas those parts of the slab which are
displaced downward will collect negative $\hat z$ momentum.  In other
words, this effect will tend to exaggerate ripples in the slab
on transverse scales greater than the slab thickness.  Clearly this effect
will have to compete with a number of stabilizing effects.  One of
the more obvious is a linear restoring force arising from the fact that
the ram pressure on the slab will be greater on surfaces that are
moving into the confining wind.  A more subtle, but more important,
effect arises from the fact that the postshock flow will tend to
converge on convex surfaces and diverge behind concave surfaces.
Consequently there is a tendency to ``fill in'' ripples in the slab
using freshly accreted material.

In what follows we will provide an analytic derivation of these effects,
and demonstrate that under appropriate circumstances a nonlinear
instability can set in, overwhelming the purely linear stabilization
of the slab.  We will start by exploring a set of model
equations, designed to capture the essential physical processes
within the slab.  Since the basic nature of this instability
is nonlinear, there is no hope of avoiding approximations.
However, we will aim to minimize their number, and to
justify them as they occur.  We will start by finding an
exact set of equations for the vertically averaged properties
of the slab, including the first order velocity moments.
Since turbulent transport processes are usually dominated
by the largest scale motions in a system, this will
allow us to arrive at a set of model equations
which include nonlinear momentum transport, while neglecting
the full range of small scale motions that actually take place
in a real shock bounded  slab.

\subsection{Formalism}

We define the boundaries of the slab by the functions $z_2(x,t)$
and $z_1(x,t)$ with $z_2>z_1$ for all $x$ and $t$.
The usual shock boundary conditions give (see appendix A)
\begin{equation}
\vec v(z_2)=\left({\dot z_2\over 1+(\partial_xz_2)^2}-
{c_s^2\over V_E+\dot z_2}\right)(\hat z-\partial_x z_2\hat x)
- {\partial_x z_2V_E\over 1+(\partial_xz_2)^2}(\hat x+\partial_x z_2\hat z),
\label{eq:vj2}
\end{equation}
\begin{equation}
\rho(z_2)=\left({V_E+\dot z_2\over c_s}\right)^2{\rho_E\over 1+(\partial_x
z_2)^2},
\label{eq:rj2}
\end{equation}
\begin{equation}
\vec v(z_1)=\left({\dot z_1\over 1+(\partial_xz_1)^2}+
{c_s^2\over V_E-\dot z_1}\right)(\hat z-\partial_x z_1\hat x)
+{\partial_x z_1V_E\over 1+(\partial_xz_1)^2}(\hat x+\partial_x z_1\hat z),
\label{eq:vj1}
\end{equation}
and
\begin{equation}
\rho(z_1)=\left({V_E-\dot z_1\over c_s}\right)^2{\rho_E\over 1+(\partial_x
z_1)^2}.
\label{eq:rj1}
\end{equation}

Next it will be useful to define the following vertically averaged
variables:

\begin{equation}
\Sigma\equiv\int_{z_1}^{z_2}\rho dz,
\label{eq:sigma}
\end{equation}
\begin{equation}
J_x\equiv\int_{z_1}^{z_2}\rho v_x dz,
\label{eq:jx}
\end{equation}
and
\begin{equation}
J_z\equiv\int_{z_1}^{z_2}\rho v_z dz.
\end{equation}
\begin{equation}
T_{ij}\equiv\int_{z_1}^{z_2}\rho v_i v_j dz.
\end{equation}

Using the mass conservation equation and Eqs. (\ref{eq:sigma}) and
(\ref{eq:jx}) we find that
\begin{equation}
\partial_t\Sigma= (\dot z_2-v_z(z_2)+\partial_x z_2 v_x(z_2))\rho(z_2)
-(\dot z_1-v_z(z_1)+\partial_x z_1 v_x(z_1))\rho(z_1)-\partial_x J_x.
\end{equation}
This can be simplified using Eqs. (\ref{eq:vj2}), (\ref{eq:rj2}),
(\ref{eq:vj1}), and (\ref{eq:rj1}) to obtain
\begin{equation}
\partial_t\Sigma=\rho_E(2V_E+\dot z_2-\dot z_1)-\partial_x J_x.
\label{eq:cont}
\end{equation}

Proceeding to the $\hat z$ component of the  momentum conservation
equation we find find that
\begin{eqnarray}
\partial_t J_z=&(\dot z_2-v_z(z_2)+\partial_xz_2
v_x(z_2))\rho(z_2)v_z(z_2)-P(z_2)
\label{eq:jz1}\\
&-(\dot z_1-v_z(z_1)+\partial_xz_1 v_x(z_1))\rho(z_1)v_z(z_1)\nonumber\\
&+P(z_1)-\partial_x T_{xz}\nonumber.
\end{eqnarray}
Using the shock jumps conditions (Eqs. (\ref{eq:vj2}), (\ref{eq:rj2}),
(\ref{eq:vj1}), and (\ref{eq:rj1})) this becomes
\begin{equation}
\partial_t J_z=-\rho_EV_E(\dot z_2+\dot z_1)-\partial_x T_{xz}.
\label{eq:jz2}
\end{equation}
Similarly we find that for $J_x$
\begin{eqnarray}
\partial_t J_x=&(\dot z_2-v_z(z_2)+\partial_xz_2 v_x(z_2))\rho(z_2)v_x(z_2)
+c_s^2\partial_xz_2\rho(z_2)
\label{eq:jx1}\\
&-(\dot z_1-v_z(z_1)+\partial_xz_1 v_x(z_1))\rho(z_1)v_x(z_1)
-c_s^2\partial_xz_1\rho(z_1)\nonumber\\
&-\partial_x T_{xx}-c_s^2\partial_x\Sigma\nonumber,
\end{eqnarray}
which becomes
\begin{equation}
\partial_t J_x=\rho_Ec_s^2(\partial_x z_2-\partial_xz_1)-\partial_x T_{xx}
-c_s^2\partial_x\Sigma.
\label{eq:jx2}
\end{equation}

In order to construct a simplified model of the slab dynamics
we need to consider the mean motion of the slab.  We can
define a mean position, $z_m$, by
\begin{equation}
z_m\equiv \Sigma^{-1}\int_{z_1}^{z_2} z\rho dz.
\label{eq:zm}
\end{equation}
This is given by
\begin{eqnarray}
\Sigma\partial_tz_m=&J_z-J_x\partial_xz_m-\partial_x\left(\int_{z_1}^{z_2}
(z-z_m)\rho v_x dz\right) +\rho_E\bigl((\dot z_2+V_E)(z_2-z_m)
\label{eq:zmt1}\\
&+(V_E-\dot z_1)(z_1-z_m)\bigr)\nonumber.
\end{eqnarray}
This can be rewritten in a slightly more convenient form as
\begin{eqnarray}
\Sigma\partial_tz_m=&J_z-J_x\partial_xz_m-\partial_x W_x
+\rho_E\bigl({1\over 2}(\dot z_2+\dot z_1)(z_2-z_1)
\label{eq:zmt2}\\
&+(2V_E+\dot z_2-\dot z_1){1\over2}(z_2+z_1-2z_m)\bigr)\nonumber.
\end{eqnarray}
where we have used
\begin{equation}
W_i\equiv \int_{z_1}^{z_2} (z-z_m)\rho v_i dz
\label{eq:jtx}
\end{equation}

By introducing $W_i$ we have committed ourselves to a consideration
of substructure within the slab.  We can't push this process
very far without ending up considering the full spectrum of small
scale motions, attempting, in other words, to construct a complete
analytic model of motions which are only partially resolved in numerical
calculations.  Still, it turns out to be critical to at least have
an approximate notion of how $W_i$ evolves.  We will assume
that we need only consider shearing motions running parallel to the
line
defined by $z_m(\vec x,t)$.  Motions normal to the slab will be assumed to
be present at whatever level is necessary to maintain quasistatic
equilibrium in response to the forces acting on the slab.  This
implies that we should define a unit vector $\hat n_m$ such that
\begin{equation}
\hat n_m\equiv {\hat x+\partial_x z_m\hat z\over (1+(\partial_x z_m)^2)^{1/2}}.
\label{eq:nmd}
\end{equation}
$\hat n_m$ defines the direction of motion parallel to the midplane
of the slab. Then we need only consider the evolution of $W_i$ in the direction
of $\hat n_m$, which we will define as $W$.
This is given by
\begin{eqnarray}
\partial_tW=&-\dot z_m \vec J\cdot \hat n_m+T_{zi}\hat n_{mi}-
\partial_x z_m T_{xi}\hat n_{mi}-\hat n_m\cdot\partial_x\left[\int_{z_1}^{z_2}
(\rho v_x\vec v+P\hat x)(z-z_m)dz\right]\\
&+\left((\dot z_2-v_z(z_2)+v_x(z_2)\partial_x z_2)\rho(z_2)\vec v(z_2)\cdot\hat
n_m
+P(\partial_xz_2\hat x-\hat z)\cdot \hat n_m\right)(z_2-z_m)\nonumber\\
&-\left((\dot z_1-v_z(z_1)+v_x(z_1)\partial_x z_1)\rho(z_1)\vec v(z_1)\cdot\hat
n_m
+P(\partial_xz_1\hat x-\hat z)\cdot \hat n_m\right)(z_1-z_m).\nonumber
\end{eqnarray}
Using Eqs. (\ref{eq:vj2}), (\ref{eq:rj2}), (\ref{eq:vj1}), and (\ref{eq:rj2})
we can rewrite this as
\begin{eqnarray}
\partial_tW=&-\dot z_m \vec J\cdot \hat n_m+T_{zi}\hat n_{mi}-
\partial_x z_m T_{xi}\hat n_{mi}-\hat n_m\cdot\partial_x\left[\int_{z_1}^{z_2}
\rho v_x\vec v(z-z_m)dz\right]\\
&-\rho_E(c_s^2+V_E^2)\hat z\cdot
n_m(z_2-z_1)-\rho_Ez_m\left(c_s^2\partial_x(z_2-z_1)
\hat x\cdot\hat n_m-V_E\partial_t(z_1+z_2)\hat z\cdot \hat
n_m\right)\nonumber\\
&+{\rho_E\over 2}\left(c_s^2\partial_x(z_2^2-z_1^2)\hat x\cdot \hat
n_m-V_E\partial_t(z_2^2+z_1^2)
\hat z\cdot\hat n_m\right).\nonumber
\label{eq:tjt}
\end{eqnarray}

Up to this point we have made no approximations.  However, our
equations are clearly simply the beginning of an infinite
hierarchy.  Moreover, the functions $z_1(\vec x,t)$ and $z_2(\vec x,t)$
are entirely unconstrained.  We need to close the hierarchy
of equations and find useful expressions for $z_1$ and $z_2$ in order
to proceed.  As long as the bulk motions within the slab are subsonic
and we are uninterested in motions with large vertical wavenumbers
we can approximate the density distribution as a smooth interpolation
between the limits given in Eqs. (\ref{eq:rj2}) and (\ref{eq:rj1}).
Then
\begin{equation}
z_m={1\over2}(z_1+z_2)+
\left({\rho(z_2)-\rho(z_1)\over\rho(z_1)+\rho(z_2)}\right){L\over 6}
\end{equation}
where $L\equiv z_2-z_1$.
If we consider the second term as a small correction it can be simplified
to
\begin{equation}
z_m\approx{1\over2}(z_1+z_2)+{L\over 6}\left({2\dot z_m\over
V_s}-\partial_xL\partial_x z_m\right)
\left(1+(\partial_x z_m)^2+{1\over 4}(\partial_xL)^2\right)^{-1}.
\label{eq:zmap1}
\end{equation}
This indicates that the difference between $z_m$ and the average of $z_1$ and
$z_2$ is
much less than $L$ unless either $\dot z_m$ is close to $V_E$ or fluctuations
in $L$
and $z_m$ are comparable to each other and to the transverse wavelength.  We
will
see that $\dot z_m$ is as large as $c_s$ only when the bending angle of the
slab
is quite large, so the correction term in Eq. (\ref{eq:zmap1}) is very small as
long as
the bending angle is small.  Unless otherwise noted we will use
\begin{equation}
z_m\approx {1\over2}(z_1+z_2)
\label{eq:zmap2}
\end{equation}
in what follows.
Consequently, we have
\begin{equation}
\Sigma={1\over 2}(\rho(z_1)+\rho(z_2))L,
\label{eq:sap}
\end{equation}
with
\begin{equation}
z_2\approx z_m+{L\over 2},
\label{eq:z1ap}
\end{equation}
and
\begin{equation}
z_1\approx z_m-{L\over 2}.
\label{eq:z2ap}
\end{equation}
We have not ruled out the possibility of a large bending angle
in these equations since the existence of a large bending
angle does not necessarily imply the failure of Eq. (\ref{eq:zmap2}).

As long as our goal is a general model of slab
behavior there seems to be little point to extending the hierarchy
of equations past the level of momentum conservation.  Higher
order moments will depend on the details of energy conservation
within the slab.  We will therefore assume that the components
of $T_{ij}$ can be given in terms of the lower order moments.
The specific form of $T_{ij}$ then follows from our general
approach.  We have chosen to consider the slab material as
though it has a net velocity (which is a function of $x$ and
$t$, but not of $z$), and a uniform gradient (which is a function
of all three coordinates, but $z$ enters only in a factor
of $z-z_m$).  Given these assumptions we have
\begin{equation}
T_{ij}\approx {J_i J_j\over\Sigma}+{12W^2\over L^2\Sigma}\hat n_i\hat
n_j
\label{eq:stress}
\end{equation}
This is equivalent to modeling the transport effects using only the
largest scale motions within the slab.  In general, there will be
smaller scale motions.  In fact, for even very small ripples
we expect the presence of a turbulent cascade within the slab.
However, even in the case of a fully developed turbulent cascade
we expect the transport properties of the medium to be dominated
by the largest scale motions (as they are for homogeneous
turbulence).  It follows that the gross dynamics of the slab
ought to be reasonably well approximated by the decomposition
given in Eq. (\ref{eq:stress}).

In a similar spirit we can approximate the integral in Eq. (\ref{eq:tjt}) as
\begin{equation}
\int_{z_1}^{z_2}\rho v_x\vec v(z-z_m)dz\approx {J_xW\over \Sigma}\hat
n_m+
{\vec JW\over \Sigma}(\hat n_m\cdot\hat x)
\label{eq:intap}
\end{equation}

Although we have arrived at a closed set of equations, there is one important
physical effect that we have neglected.  The existence of a nonzero $W$
implies a strong shear within the slab.  In an inviscid fluid we expect this
to lead to
the existence of a local Kelvin-Helmholtz instability.  We can express
the criterion for the onset of such an instability by considering the
shear caused by the existence of a thin postshock layer with a velocity
$V_o$.  If the shock
surface changes its properties on a time scale $\tau$ then the local shear
will be $(V_o t/L\tau)$.  This will lead to turbulent mixing
with the adjacent layers if this shear times the time available before the
layer collides with a similar layer generated at a distance $>k^{-1}$ is
large.  In other words, we expect to find turbulent mixing if
\label{eq:khc}
\begin{equation}
\left[{V_o t\over L\tau}\right]{1\over kV_o}\sim {1\over kL}{t\over\tau}\gg 1.
\end{equation}
If the dynamical time scale is very long, i.e. comparable to $t$, then this
implies
that turbulent mixing will still be at least marginally effective for the high
velocity postshock layer.
We will see in what follows that typical rates for both bending and breathing
modes are of order $c_sk$, and never less than $c_s k (Lk)^{1/2}$, so
that for $kL<1$ the condition for turbulent mixing is always satisfied by a
factor of at least $V_E/c_s$. The resulting turbulence
will cause an averaging of the momentum within the disk (which helps make
our vertically averaged model more plausible), but also gives rise to a
damping of $W$
at a rate given roughly by the velocity gradient across the slab.  We can
insert this effect into Eq. (\ref{eq:tjt}) by introducing a damping term
equal to
\begin{equation}
-C_d \left|{12W\over L^2\Sigma}\right|(1+(\partial_xz_m)^2)^{1/2}
W,
\label{eq:jtd}
\end{equation}
where $C_d$ is a constant of order unity related to the angle of turbulent
entrainment and the rest of the coefficient of $W$ is just the
gradient of the velocity across the slab, in the direction normal to the
slab.  Once again we note that this choice follows from our assumption
that the transport properties of any motions within the slab
are approximately those expected from homogeneous turbulence,
and are therefore dominated by the largest scale motions within
the disk, in this case eddys with a shear of roughly $W/(\Sigma L^2)$
and a scale of $L$.

Eqs. (\ref{eq:rj1}), (\ref{eq:rj2}), (\ref{eq:cont}), (\ref{eq:jz2}),
(\ref{eq:jx2}), (\ref{eq:zmt2}), (\ref{eq:nmd}), (\ref{eq:tjt}),
(\ref{eq:zmap1}),
(\ref{eq:sap}), (\ref{eq:z1ap}), (\ref{eq:z2ap}), (\ref{eq:stress}),
(\ref{eq:intap}) and
(\ref{eq:jtd}), form a complete set of equations giving an approximate
description of the evolution of a shock bounded slab.  Unfortunately, in this
form their implications are unclear.  In what follows we will find approximate
forms for different regimes, in particular the case where $|\partial_xz_m|$ is
small, but nonlinear transport effects can produce a bending instability.

\subsection{Linear Theory}

We begin with the linear theory of breathing modes of the slab, i.e.
fluctuations
in the slab column density.  These turn out to be relevant to the theory of
gravitational instabilities in slabs, although not to the instability which
is the main focus of this paper.  To linear order Eq. (\ref{eq:cont}) becomes
\begin{equation}
\dot{\delta\Sigma}-\rho_E\dot{\delta L} =-\partial_x J_x.
\label{eq:lbr1}
\end{equation}
At this level of approximation all the components of $T_{ij}$ are zero so
Eq. (\ref{eq:jx2}) is just
\begin{equation}
\dot J_x =\rho_Ec_s^2\partial_x L-c_s^2\partial_x\Sigma.
\label{eq:lbr2}
\end{equation}
This set of equations can be closed using Eqs. (\ref{eq:rj1}), (\ref{eq:rj2}),
(\ref{eq:sap}).  We find that
\begin{equation}
\delta\Sigma={\Sigma\over L}\left(\delta L+\left[{L\dot{\delta L}\over V_s
(1+2/D)}\right]\right),
\label{eq:lbr3}
\end{equation}
where $D\equiv V_s^2/c_s^2$ is the density ratio across the shock face.  We
will assume that $D\gg 1$
Eqs. (\ref{eq:lbr1}), (\ref{eq:lbr2}), and (\ref{eq:lbr3}) can be combined
into a single linear equations for $\delta L$.   We find that
\begin{equation}
\ddot{\delta L}\left(1+{1\over D}\right)+{2\over D}\partial_t(t\ddot{\delta
L})=
-c_s^2k^2\left(\delta L\left(1-{1\over D}\right)+{2t\over D}\dot{\delta
L}\right).
\end{equation}
Retaining terms to first order in $D^{-1}$ we find that
\begin{equation}
\delta L\propto e^{ikx} e^{i\omega t}
\end{equation}
where
\begin{equation}
\omega=\pm c_sk \left(1-{2\over D}\right).
\label{eq:lres1}
\end{equation}
We see that the shock bounded slab is stiff in the sense that in the limit
where
$D\rightarrow\infty$ it resists longitudinal compression just as strongly as
if the slab material were confined within two rigid boundaries.

We now turn to the bending modes of the slab, which will turn out to be
unstable in the nonlinear regime.  In this limit Eqs. (\ref{eq:jz2}),
(\ref{eq:zmap2}),
and (\ref{eq:zmt2}) become
\begin{equation}
\dot J_z=-\rho_EV_E(\dot z_1+\dot z_2),
\label{eq:lb1}
\end{equation}
and
\begin{equation}
\Sigma\dot z_m=J_z-\partial_xW_x+\rho_EL\dot z_m +V_s(z_2+z_1-2z_m).
\label{eq:lb2}
\end{equation}
Taking the derivative of Eq. (\ref{eq:lb2}) and using Eq. (\ref{eq:lb1})
we find that
\begin{equation}
\left(1-{1\over D}\right)\Sigma\left[{1\over t}\dot z_m+\ddot z_m\right]=
(V_s-V_E)\rho_E2\dot z_m-2\dot z_m V_s\rho_E-ik\partial_tW_x
\label{eq:lb4}
\end{equation}
Next we note that to linear order $W_x=W$.  Combining
Eqs. (\ref{eq:nmd}), (\ref{eq:tjt}), and (\ref{eq:zmap2}) we obtain
\begin{equation}
\partial_tW=-\rho_EV_E^2 L\partial_x z_m \left(1+{1\over D}\right),
\end{equation}
or
\begin{equation}
\partial_tW=-\Sigma c_s^2\partial_x z_m\left(1-{1\over D}\right).
\label{eq:lb3}
\end{equation}
Combining Eqs. (\ref{eq:lb4}) and (\ref{eq:lb3}) we get
\begin{equation}
\ddot z_m+{2\over t}\dot z_m+k^2c_s^2z_m=0,
\end{equation}
which has solutions of the form
\begin{equation}
z_m\propto{1\over t}e^{i(kx\pm c_skt)}.
\label{eq:lres2}
\end{equation}
We note that $\Sigma |\dot z_m|\gg J_z$.  The slab is not moving back
and forth very much.  Instead, the ripples are being erased by the preferential
deposition of fresh material in concave regions.

Eqs. (\ref{eq:lres1}) and (\ref{eq:lres2}) indicate that a shock bounded slab
is quite stiff.  By contrast, a pressure bounded slab is quite pliable.
One might be tempted to ascribe this result to the approximations used, but
in Appendix B we show that the same result follows from a linear perturbation
calculation that avoids vertical averaging.  In fact, Eq. (\ref{eq:lres1}) is
in complete agreement with the more exact treatment in appendix B.  Eq.
(\ref{eq:lres2})
is not, except in the limit where $D\rightarrow\infty$.  Evidently vertical
averaging does exact some penalty.  Still, for the interesting case where $D$
is large the difference is only a small correction to the oscillation
frequency.
We conclude that the detailed structure of the postshock flow is largely
irrelevant
to the averaged behavior of the slab.  This is not actually very surprising,
since
the behavior of the vertically averaged horizontal mass flux is the most
important part
of the slab dynamics.  Conservation of momentum guarantees that this quantity
is insensitive to whether or not the immediate postshock layer undergoes
turbulent
mixing.  The only loophole is the possibility that the postshock layer might
dissipate by colliding with an opposing flow, generated at a place where the
slab
curvature is significantly different.  For reasons noted above, this is
physically implausible.

The stiffness of a shock-bounded slab has been noticed before
(\cite{v83,vo88}) in the context of breathing modes.  The fact that
bending modes are also stiff is a new result.  In either case this shows that
serious errors can be introduced whenever the shock boundary conditions are
approximated by an exterior pressure (e.g. \cite{ee78,ws81,lp93}).

\subsection{Nonlinear Theory in the Small Bending Angle Limit}

At what point does the linear analysis become inadequate?  This will
clearly be the case when the bending angles, $kz_m$ or $k\delta L$,
are of order unity.  On the other hand, when these angles are small
nonlinear effects may still be important.  This turns out to be
particularly true for bending modes.  In this section we will
show this by rewriting our basic equations in the limit where the bending
angles
are small and $D\gg 1$.

Eq. (\ref{eq:cont}) becomes
\begin{equation}
\dot {\delta\Sigma}=-\partial_x J_x,
\label{eq:br1}
\end{equation}
since the first term on the right hand side of Eq. (\ref{eq:cont})
is of order $D^{-1}$ relative to the left hand side.
Eqs. (\ref{eq:jz2}), (\ref{eq:zmap2}), (\ref{eq:nmd}), and (\ref{eq:stress})
can be combined as
\begin{equation}
\dot J_z=-\partial_x\left({J_xJ_z\over\Sigma}+{12W^2\partial_x z_m\over
L^2\Sigma}\right)
-\rho_E2V_E\dot z_m.
\label{eq:be1}
\end{equation}
The last term on the right hand side dominates in the linear approximation, but
is
dynamically unimportant since $J_z\ll\dot z_m$ in this limit.
Similarly Eqs.  (\ref{eq:jx2}), (\ref{eq:zmap2}), (\ref{eq:nmd}), and
(\ref{eq:stress})
becomes
\begin{equation}
\dot J_x=-c_s^2\partial_x\Sigma-\partial_x\left({J_x^2\over\Sigma}+{12
W^2\over L^2\Sigma}\right).
\label{eq:br2}
\end{equation}
In the same limit Eq. (\ref{eq:zmt2}) becomes
\begin{equation}
\Sigma\dot z_m=J_z-J_x\partial_xz_m-\partial_xW.
\label{eq:be2}
\end{equation}
Next, Eqs. (\ref{eq:nmd}), (\ref{eq:tjt}), (\ref{eq:stress}), (\ref{eq:intap})
and (\ref{eq:jtd}) imply
\begin{eqnarray}
\partial_tW=&-\dot z_m\vec J\cdot(\hat x+\partial_x z_m\hat z)
+{1\over\Sigma}\left(\partial_x z_m(J_z^2-J_x^2)+J_xJ_z\right)
-\partial_x\left[{2W J_x\over\Sigma}\right]
\label{eq:be3}\\
&-\partial_xz_m\partial_x\left[{J_zW\over\Sigma}\right]
-\rho_EV_E^2\partial_x z_m L-C_d\left|{12W\over L^2\Sigma}\right|
W.
\nonumber
\end{eqnarray}
Using Eq. (\ref{eq:be2}) we can simplify this somewhat and obtain
\begin{equation}
\partial_tW=-W\partial_x\left({J_z\over\Sigma}\right)\partial_xz_m+
{J_x\over\Sigma}\partial_xW -\partial_x\left[{2W J_x\over\Sigma}\right]
-\Sigma c_s^2\partial_xz_m-C_d\left|{12W\over L^2\Sigma}\right|W
\label{eq:wnu}
\end{equation}

Finally, we note that to the accuracy required here we can take
\begin{equation}
{\delta\Sigma\over\Sigma}={\delta L\over L}
\end{equation}

We are now in a position to see when the linear approximation becomes
inadequate.
For breathing modes we know that to linear order $J_x\sim c_s\delta\Sigma$.
{}From symmetry considerations (as well from inspection of the preceding
paragraphs)
there is no obvious coupling to bending modes even when the breathing modes
become quite strong.  This means that we can restrict our attention to Eqs.
(\ref{eq:br1}) and (\ref{eq:br2}).  The only nonlinear term is in Eq.
(\ref{eq:br2})
and we can see that it becomes important when $\delta\Sigma\sim\Sigma$.  In
other
words, a linear treatment of these modes fails when the slab velocities are of
order the sound speed and the column density fluctuates by factors of order
unity.
This is the expected result given that the breathing modes of a slab behave
like sound waves in a shock tube.  In any case, once the bulk velocities are of
order the sound speed it is unlikely that a quasistatic treatment of the
internal structure of a slab will shed any light on the actual dynamics.

For the bending modes, in the linear limit, we have $J_z\sim \Sigma z_m/t$,
and $W\sim c_s\Sigma z_m$.  In this case we find that there are
nonlinear
effects which will give $J_x\sim c_s\Sigma (z_m/L)^2$ and $\delta\Sigma\sim
\Sigma(z_m/L)^2$.
Furthermore, the nonlinear terms in Eq. (\ref{eq:be1}) will dominate as soon
as $z_m$ is of order $L/D^{1/2}$.  In spite of this the nonlinear terms
in Eq. (\ref{eq:be2}) do not contribute significantly to the evolution of $z_m$
until $z_m\sim L$.  On the other hand, the damping term in Eq. (\ref{eq:wnu})
dominates secular damping when $z_m>L^2/(c_st)$ and produces strong
damping when $z_m$ is of order $kL^2$.  None of the other nonlinear
terms in Eq. (\ref{eq:wnu}) are important unless $z_m\sim L$.  This suggests
a simple set of model equations for bending modes, applicable when typical
values of $|z_m|$ are less than $L$ and $kL<1$.
{}From Eqs. (\ref{eq:be2}) and (\ref{eq:wnu})
we get
\begin{equation}
\Sigma\dot z_m\approx -\partial_xW,
\end{equation}
and
\begin{equation}
\partial_tW=\Sigma c_s^2\partial_x z_m-C_d\left|{12W\over
L^2\Sigma}\right|W.
\end{equation}
Our failure to include $J_z$ in the dynamics means that we have sacrificed
our ability to follow secular trends in the mode amplitude.  It follows that
we can go ahead and neglect the time dependence of $\Sigma$ without further
loss of accuracy.  Combining our model equations we find
\begin{equation}
\partial_t^2W=c_s^2\partial_x^2W-{24C_d\over L^2\Sigma}|
W|\partial_tW.
\label{eq:damp1}
\end{equation}
In other words, the turbulent damping term provides an amplitude dependent
friction
term.  When this term is important, i.e. when the damping rate can be greater
than
$c_sk$, the effect is to cause $\partial_tW$ to quickly converge to
\begin{equation}
\partial_tW\approx -{L^2\Sigma k^2c_s^2\over 24 C_d}\hbox{ sign}(
W),
\label{eq:overd}
\end{equation}
so that $W$ moves toward zero with a speed which does not depend on its
amplitude
(as long as $|W|$ is large enough for the damping term to dominate the
dynamics).
At these large amplitudes the rate at which the bending mode evolves is
of order
\begin{equation}
\tau^{-1}\sim \left({kL^2\over C_d <z_m^2>^{1/2}}\right)^{1/2} kc_s.
\end{equation}
Here $<z_m^2>^{1/2}$ refers to a root mean square value of $z_m$ after
averaging
over all phases of the oscillation.  Hereafter we will drop these brackets
for order of magnitude estimates.  We see that turbulent damping slows
the bending oscillations.  In that sense the slab is less stiff for strong
bending
motions than for weak ones, although it would be more accurate to
say that in this limit the bending modes are overdamped.
As long as $z_m$ lies between $L$ and
$kL^2$ we can estimate the order of the dynamical quantities.  We have
\begin{equation}
W\sim \Sigma c_s z_m C_d^{-1/2}\left({kL^2\over z_m}\right)^{1/2},
\end{equation}
\begin{equation}
J_x\sim \Sigma c_s kL C_d^{-3/2}\left(kz_m\right)^{1/2},
\end{equation}
\begin{equation}
J_z\sim \Sigma c_s \left({z_m\over L}\right)\left(kz_m\right)^{3/2} C_d^{-1/2},
\end{equation}
and
\begin{equation}
\delta\Sigma\sim \Sigma C_d^{-1} kz_m.
\end{equation}
We note that in this regime the fluctuations in $\Sigma$ are just
large enough to suppress the terms driving $J_x$.  In other words,
the slab column density and transverse momentum evolve quasistatically.
In this case, as for smaller amplitudes, $|\Sigma\dot z_m|\gg J_z$.
The rippling of the slab is still dominated by differential deposition
of material in the postshock layer rather than bulk motion of the slab.
The actual motions in the slab (roughly $\vec J/\Sigma$) are subsonic
even as $<z_m^2>^{1/2}\rightarrow L$.

What happens when $z_m>L$?  This is consistent with the weak bending
limit only for $kL<1$.  In this limit we can get a consistent ordering
of the dynamical quantities in Eqs. (\ref{eq:br1}) through (\ref{eq:wnu}).
We get a dynamical rate, $\tau^{-1}$, given by
\begin{equation}
\tau^{-1}\sim c_sk kz_m C_d^{-1/2}.
\label{eq:z4}
\end{equation}
The other dynamical variables are of order
\begin{equation}
W\sim \Sigma c_s L kz_m C_d^{-1/2},
\label{eq:zn1}
\end{equation}

\begin{equation}
J_z\sim \Sigma c_s (kz_m)^{3/2} C_d^{-1/2},
\end{equation}

\begin{equation}
J_x\sim \Sigma c_s kz_m C_d^{-3/2},
\end{equation}
and
\begin{equation}
\delta\Sigma\sim \Sigma kz_m C_d^{-1}.
\label{eq:zn2}
\end{equation}
With this ordering the dynamical Eqs. (\ref{eq:br1}) through
(\ref{eq:wnu}) become
\begin{equation}
\partial_t\delta\Sigma\approx -\partial_x J_x,
\end{equation}

\begin{equation}
\dot J_z\approx -\partial_x\left({12W^2\partial_xz_m\over
L^2\Sigma}\right),
\label{eq:z1}
\end{equation}

\begin{equation}
c_s^2\partial_x\Sigma\approx -\partial_x\left({12W^2\over
L^2\Sigma}\right),
\end{equation}

\begin{equation}
\Sigma\dot z_m\approx J_z,
\label{eq:z2}
\end{equation}
and
\begin{equation}
\Sigma c_s^2\partial_x z_m\approx -C_d{12\over L^2\Sigma}
|W|W.
\label{eq:z3}
\end{equation}

Note that in this limit the bending modes of the slab have
become dominated by the bulk motion in the $\hat z$ direction.
Since $|\delta\Sigma|<\Sigma$ we can combine Eqs. (\ref{eq:z1}),
(\ref{eq:z2}), and (\ref{eq:z3}) to obtain
\begin{equation}
\ddot z_m\approx -\partial_x\left({(\partial_xz_m)^2 c_s^2
\hbox{ sign}(\partial_x z_m)\over C_d}\right).
\label{eq:star}
\end{equation}
We can see that this implies an instability by assuming an initial $z_m$
of $A\sin(kx)$.  Eq. (\ref{eq:star}) implies
\begin{equation}
\ddot A\approx {c_s^2k^3 A^2\over C_d}\sin(2kx)\hbox{ sign}(\cos(kx)),
\end{equation}
which has a component proportional to $\sin(kx)$ with a coefficient
of $8c_s^2k^3A^2/(3\pi C_d)$.  Evidently ripples with $z_m$ at least
of order $L$ are driven to larger amplitudes at a rate given approximately
by Eq. (\ref{eq:z4}).  This is the Nonlinear Thin Shell Instability (NTSI).

The physical mechanism driving the NTSI is just the net transport of positive
(negative) $\hat z$ momentum toward parts of the slab displaced in the
positive (negative) $\hat z$ direction by the anti-symmetric  flow with the
slab ($W$), which is in turn driven by the obliquity of the shock
fronts when the slab is bent.  Since the driving mechanism is inherently
nonlinear it will tend to couple motions on different length scales.
However, an examination of this coupling really requires Eqs. (\ref{eq:br1})
through (\ref{eq:wnu}) rather than Eqs. (\ref{eq:z1}) through (\ref{eq:z3})
since when $\partial_xz_m$ is small, as at local extrema of the slab
displacement,
Eq. (\ref{eq:z3}) gives $W\approx 0$, which is unrealistic.  In other
words, Eqs. (\ref{eq:z1}), (\ref{eq:z2}), and (\ref{eq:z3}), while
giving a good estimate of the average dynamics of the slab, will underestimate
the acceleration of the extrema.

Our whole approach will break down as $\partial_x z_m$ goes to $1$.  In this
limit
both components of $\vec J$ go to $\Sigma c_s$, and $|W|\rightarrow
\Sigma c_sL$,
i.e. the bulk flow of the slab and the internal motions within it will no
longer be
subsonic.  As this instability grows the slab will tend to collect mass at its
extrema.  The fractional mass fluctuations will approach unity as
$kz_m\rightarrow 1$.
One might be tempted to regard this as the fragmentation of the slab, given the
large mass fluctuations and bulk velocities.  In a sense this is true, but the
available simulations (\cite{ste92}) show that slab still exists as a
continuous thin region of high density.

The requirement that the slab must be displaced by at least a distance
comparable
to $L$ would appear, at first glance, to imply that this instability is not
easy to excite.  This will certainly be true when $D$ is not very large.
On the other hand, for $D$ large a displacement of order $L$ is almost
inevitable,
which is why this instability showed up in simulations of colliding flows
with only numerical noise for a seed (\cite{hun86,ste92}).

Once the NTSI shows up it will grow to saturation at a rate $>c_sk(kL)^{1/2}$,
which will be much faster than $t^{-1}$ for a broad range of $k$.  For $k$
so small that $c_sk$ is of order $t^{-1}$ or less the NTSI will be suppressed
by secular damping terms.  In fact,
if we define $R$ as the propagation distance over which the zeroth order shock
properties evolve (equal to $V_Et$ for a stationary slab)  then we see that
the NTSI exists in the range
\begin{equation}
L^{-1}\gg k\gg \left({V_s\over c_s}\right) R^{-1}.
\end{equation}
This suggests another reason why a high $D$ is necessary to observe the
NTSI.  In general this range will have a dimensionless width proportional
to $D^{1/2}$.  For small $D$ the range of unstable $k$ can disappear.

There is one remaining loophole in our treatment here.  We have
assumed throughout that the postshock layer undergoes turbulent
mixing before traveling a distance comparable to the perturbation
wavelength.  Although our estimate for the onset of a Kelvin-Helmholtz
instability suggests that we are on safe ground here, it is still
interesting to check the sensitivity of our results to this assumption.
Suppose we imagine that the postshock flow, with a velocity of
order $V_0\sim V_s kz_m$,
is confined to a thin layer composed of freshly accreted material that
has not yet traveled a substantial fraction of a perturbation wavelength.
This implies a postshock depth of $L(kV_0t)^{-1}$ (assuming that the
travel time $(kV_0)^{-1}$ is less than the perturbation dynamical time)
and a bulk slab acceleration of
\begin{equation}
\Sigma \ddot z_m \sim k\Sigma (V_0^2kz_m) (kV_0t)^{-1},
\end{equation}
implying an instability growth rate of
\begin{equation}
\tau^{-1}\sim c_s k\left({z_m\over L}\right)^{1/2}.
\label{eq:g2}
\end{equation}
The condition that $\tau{-1}<kV_0$ is then just
\begin{equation}
kz_m>{c_s\over V_s}.
\label{eq:condition}
\end{equation}
In order for the NTSI to appear $\tau^{-1}$ has to exceed the rate
at which differential filling of concavities and convexities will
erase the ripples.  In this limit, and assuming Eq. (\ref{eq:condition})
is satisfied, this is
\begin{equation}
\tau^{-1}>{L\over z_m t},
\end{equation}
or
\begin{equation}
z_m>L\left({c_s\over V_s}\right)^{2/3} (kL)^{-2/3}.
\label{eq:onset}
\end{equation}
Since $c_skt>1$ we see that Eq. (\ref{eq:onset}) is easier to
satisfy than $z_m>L$ and  Eq. (\ref{eq:g2}) gives a faster
growth rate than Eq. (\ref{eq:z4}).  In the somewhat improbable
limit where the postshock flow is confined to a thin layer
(in spite of being significantly supersonic at the onset of the
instability) the NTSI will be easier to excite and will grow
at a faster rate.

\section{Some Asymmetric Slabs}

The physical mechanism behind the NTSI is quite general.
Does it show up in other situations?  In this section we
will consider two other cases involving shocks with strong
compression ratios.  The first is a stationary slab, similar
to the one considered in the previous section, but with a single
confining shock.  The opposing surface will be assumed to
be confined by a infinitely thin, infinitely hot gas.  We
retain the assumption that the slab is neither accelerating
nor decelerating since in those cases the slab is known
to be linearly unstable.  The second case is a decelerating,
highly radiative blast wave with no significant thermal pressure
behind the cold postshock layer.

\subsection{The Stationary Asymmetric Slab}

If $z_1$ defines a contact discontinuity then Eqs. (\ref{eq:vj1})
and (\ref{eq:rj1}) are replaced by
\begin{equation}
\vec\nabla\cdot\vec v(z_1)=0,
\end{equation}
\begin{equation}
\dot z_1-v_z(z_1)+v_x(z_1)\partial_x z_1=0,
\end{equation}
and
\begin{equation}
\rho(z_1)=\rho_E{V_s^2\over c_s^2}.
\end{equation}

The dynamical equations are then
\begin{equation}
\dot \Sigma=\rho_E(V_E+\dot z_2)-\partial_x J_x,
\label{eq:sl1}
\end{equation}
\begin{equation}
\dot J_x=\rho_Ec_s^2\partial_x z_2-\rho_EV_s^2\partial_xz_1
-\partial_xT_{xx}-c_s^2\partial_x\Sigma,
\label{eq:sl2}
\end{equation}
\begin{equation}
\dot J_z=-\rho_EV_E\left(\dot z_2-{c_s^2\over V_s}\right)-\partial_xT_{xz},
\label{eq:sl3}
\end{equation}
\begin{equation}
\Sigma\dot z_m=J_z-J_x\partial_xz_m-\partial_xW_x+\rho_E
(\dot z_2+V_E)(z_2-z_m),
\label{eq:sl4}
\end{equation}
and
\begin{eqnarray}
\partial_tW=&-\dot z_m \vec J\cdot \hat n_m+T_{zi}\hat n_{mi}-
\partial_x z_m T_{xi}\hat n_{mi}-\hat n_m\cdot\partial_x\left[\int_{z_1}^{z_2}
\rho v_x\vec v(z-z_m)dz\right]
\label{eq:sl5}\\
&-\rho_E(z_2-z_m)\left((V_E(V_E+\dot z_2)+c_s^2)\hat z-\partial_xz_2c_s^2\hat
x\right)
\cdot\hat n_m\nonumber\\
&-\rho_E V_s^2(z_1-z_m)(\partial_xz_1\hat x-\hat z)\cdot\hat n_m.
-C_d \left|{12W\over L^2\Sigma}\right|(1+(\partial_xz_m)^2)^{1/2}
W,
\nonumber
\end{eqnarray}
If we take $z_2\approx z_m+L/2$ and $\Sigma\propto L$ and take the
small bending limit then Eqs. (\ref{eq:sl1}),
(\ref{eq:sl2}), (\ref{eq:sl4}), and (\ref{eq:sl5}) can be rewritten as
\begin{equation}
\dot L{\Sigma\over L}=\rho_E(V_E+\dot z_2)-\partial_x J_x,
\label{eq:sl1a}
\end{equation}
\begin{equation}
\dot J_x=-c_s^2\rho_E(D-1)\partial_xz_2 -\partial_xT_{xx},
\label{eq:sl2a}
\end{equation}
\begin{equation}
\Sigma\dot z_2=J_z-J_x\partial_xz_2-\partial_xW_x
+\rho_EL(\dot z_2+V_E)+{1\over 2}(J_x\partial_xL-L\partial_xJ_x),
\label{eq:sl4a}
\end{equation}
and
\begin{eqnarray}
\partial_tW=&-\dot z_m \vec J\cdot \hat n_m+T_{zi}\hat n_{mi}-
\partial_x z_m T_{xi}\hat n_{mi}-\hat n_m\cdot\partial_x\left[\int_{z_1}^{z_2}
\rho v_x\vec v(z-z_m)dz\right]
\label{eq:sl5a}\\
&-\rho_E{L\over 2}V_sV_E\partial_xz_2
-C_d \left|{12W\over L^2\Sigma}\right|W.
\nonumber
\end{eqnarray}
In the last equation we also assumed that $\dot z_2$ can be ignored compared
to $V_s$, an approximation which is consistent with taking a uniform density
in the perturbed slab. Eqs. (\ref{eq:sl1a}) through (\ref{eq:sl5a})
can be closed by invoking Eqs. (\ref{eq:stress}) and (\ref{eq:intap}).

We can see immediately that Eqs. (\ref{eq:sl3}), (\ref{eq:sl1a}),
(\ref{eq:sl2a}),
(\ref{eq:sl4a}), and (\ref{eq:sl5a}) admit a trivial solution consisting of
any static $\delta L$ with all other dynamic variables equal to zero.  This is
just a restatement of the obvious point that perturbations in the pressure
bounded
face have no dynamical consequences.  If we linearize these equations and
discard terms of order $D^{-1}$ we find that
\begin{equation}
\partial_t^2\delta z_2+{2\over t}\partial_t\delta z_2=c_s^2\partial_x^2z_2,
\label{eq:sll1}
\end{equation}
where $\delta z_2\equiv \dot z_2-c_s^2/V_s$.  Eq. (\ref{eq:sll1}) implies that
\begin{equation}
z_2\propto {1\over t} e^{i\pm c_sk t}.
\end{equation}
The other dynamical variables are
\begin{equation}
J_z={\Sigma\over t}\delta z_2,
\label{eq:ssjz1}
\end{equation}
\begin{equation}
J_x=\mp\Sigma c_s{\delta z_2\over L},
\end{equation}
\begin{equation}
W=\mp{1\over 2}\Sigma c_s\delta z_2,
\end{equation}
and
\begin{equation}
\delta L=\delta z_2.
\end{equation}
In this case the division between bending and breathing modes is no longer
useful.
There is only one set of nontrivial modes, and they involve transverse motions,
column density fluctuations, and internal shear.  The frequencies of the linear
modes are just $\omega^2=0, c_s^2k^2$, in agreement with the detailed linear
theory given in appendix C.  The above estimate for $J_z$
fails when the perturbations are only slightly nonlinear and should be
replaced by
\begin{equation}
J_z\sim \Sigma c_s kL \left({\delta z_2\over L}\right)^3,
\end{equation}
which is the order of the nonlinear contribution from the $W^2$
part of $T_{xz}$.  This increase in $J_z$ does not affect the dynamics
of the system.

As before, we can see that as the amplitude of $\delta z_2$ is increased the
first nonlinear term to play an important role is the nonlinear damping term
in Eq. (\ref{eq:sl5a}).  However, in this case its role is quite different.
Once $z_2$ is of order $kL^2/C_d$ or greater the evolution of $W$
is strongly affected by the damping term, but this does not lead to a
slower evolutionary time scale for the perturbations.  If we write
down the equation analogous to Eq. (\ref{eq:damp1}) we have
\begin{equation}
\partial_t^3W=c_s^2\partial_x^2\partial_tW+c_s^2\partial_x^2
\left({6C_d\over \Sigma L}|W|W\right)-{12C_d\over \Sigma L^2}
\hbox{ sign}(W)\partial_t^2W^2.
\label{eq:sta1}
\end{equation}
We can immediately see that when the damping term is important a third
solution appears with an inverse time scale like the damping rate (which is,
by assumption, larger than $c_sk$).  However, the remaining solutions
still have rates of order $c_sk$.  In fact, if we approximate Eq.
(\ref{eq:sta1})
assuming we are interested only in such solutions we find that
\begin{equation}
\partial_t^2W^2={c_s^2\over 2}\partial_x^2W^2.
\end{equation}
This can be recast in a simpler form if we note that for these solutions
\begin{equation}
W^2\approx {(\Sigma c_s L)^2\over 24 C_d} |\partial_x z_2|.
\end{equation}
To leading order these solutions are oscillations with a frequency which
is smaller than the linear result by a factor of $2^{-1/2}$.
Consequently, in this regime, i.e. when $kL^2<\delta z_2< L$, we have
oscillations with
\begin{equation}
\tau^{-1}\sim c_sk
\end{equation}
\begin{equation}
\delta L\sim \delta z_2,
\end{equation}
\begin{equation}
J_x\sim \Sigma c_s {\delta z_2\over L},
\end{equation}
\begin{equation}
J_z\sim \Sigma c_s (k\delta z_2)^2 C_d^{-1},
\end{equation}
and
\begin{equation}
W\sim \Sigma c_s L (k\delta z_2)^{1/2} C_d^{-1}.
\end{equation}

The fact that the overdamped regime does not exist in this case eliminates
the possibility of the NTSI.  We can see, for example, that as
$\delta z_2\rightarrow L$, $J_x\rightarrow \Sigma c_s$.  Consequently, as the
ripples in the shock front become large the bulk flow in the slab becomes
supersonic, eliminating the justification for assuming that the slab
is in approximate hydrostatic equilibrium.  Even putting this to one side,
we see that
\begin{equation}
{\Sigma\partial_t\delta z_2\over J_z}\sim {1\over k\delta z_2}.
\end{equation}
The bulk motion in the $\hat z$ direction has little affect on the ripples in
the shock
front as long as we are in the small bending limit.  This eliminates the
possibility
of driving the NTSI, which depends on the nonlinear transport of $\hat z$
momentum
within the slab and its subsequent affect on the slab geometry.
The large amplitude limit of slab oscillations in this case has more in common
with
the nonlinear limit of the breathing mode in the symmetric case than the
bending
mode, in spite of the fact that $W\ne 0$ in this case.

In spite of these evident difficulties we can try estimating the order of the
dynamical quantities in the limit where $z_2>L$.  The evolution rate is
no longer $c_sk$, but instead the advective terms involving $J_x$ dominate
and $\tau^{-1}\sim kJ_x/\Sigma$.  Once again invoking Eqs. (\ref{eq:sl3}),
(\ref{eq:sl1a}), (\ref{eq:sl2a}), (\ref{eq:sl4a}), and (\ref{eq:sl5a}) we
estimate the relative magnitudes of the dynamic variables as
\begin{equation}
\tau^{-1}\sim c_sk \left({\delta z_2\over L}\right)^{1/2}
\end{equation}
\begin{equation}
\delta L\sim \delta z_2,
\end{equation}
\begin{equation}
J_x\sim \Sigma c_s \left({\delta z_2\over L}\right)^{1/2},
\end{equation}
\begin{equation}
J_z\sim \Sigma c_s (k\delta z_2)^2 \left({L\over\delta
z_2}\right)^{1/2}C_d^{-1},
\end{equation}
and
\begin{equation}
W\sim \Sigma c_s L (k\delta z_2)^{1/2} C_d^{-1}.
\end{equation}
These results cannot establish stability for $\delta z_2$ large,
since the dynamics are complicated and we have not attempted to solve
them, but we can see that this ordering of the dynamical quantities is
inconsistent with the NTSI.  In particular we
note that for $kL\ll1$ the bulk motion of the slab never dominates
$\partial_t\delta z_2$.

\subsection{The Cold Decelerating Slab}

If we consider a decelerating slab bounded on one side by a shock, then the
postshock
density profile is an exponential with a scale height $L_s=c_s^2/|\dot V|$,
where
$|\dot V|$ is the net deceleration of the slab.  The boundary conditions at
$z_1$
are replaced by the requirement that $\rho\rightarrow 0$ far behind the shock
front.
In this case the dynamical equations become
\begin{equation}
\dot \Sigma=\rho_E(V_E+\dot z_2)-\partial_x J_x,
\label{eq:sld1}
\end{equation}
\begin{equation}
\dot J_x=\rho_Ec_s^2\partial_x z_2-\partial_xT_{xx}-c_s^2\partial_x\Sigma,
\label{eq:sld2}
\end{equation}
\begin{equation}
\dot J_z=-\rho_EV_s^2-\rho_EV_E\left(\dot z_2-{c_s^2\over
V_s}\right)-\partial_xT_{xz},
\label{eq:sld3}
\end{equation}
\begin{equation}
\Sigma\dot z_m=J_z-J_x\partial_xz_m-\partial_xW_x+\rho_E
(\dot z_2+V_E)(z_2-z_m),
\label{eq:sld4}
\end{equation}
\begin{equation}
L_s=z_2-z_m,
\label{eq:lsca}
\end{equation}
and
\begin{eqnarray}
\partial_tW=&-\dot z_m \vec J\cdot \hat n_m+T_{zi}\hat n_{mi}-
\partial_x z_m T_{xi}\hat n_{mi}-\hat n_m\cdot\partial_x\left[\int_{z_1}^{z_2}
\rho v_x\vec v(z-z_m)dz\right]
\label{eq:sld5}\\
&-\rho_E(z_2-z_m)\left((V_E(V_E+\dot z_2)+c_s^2)\hat z-\partial_xz_2c_s^2\hat
x\right)
\cdot\hat n_m\nonumber\\
&-C_d \left|{W\over
2L_s^2\Sigma}\right|(1+(\partial_xz_m)^2)^{1/2}W.
\nonumber
\end{eqnarray}
Here
the velocity $\dot z_2$ is defined in the frame in which the unperturbed
postshock gas is instantaneously at rest.
The coefficient of the damping term in Eq. (\ref{eq:sld5}) is necessarily a bit
more ambiguous than in our previous cases where we could estimate it by
assuming
a uniform velocity gradient across a slab with uniform density.  In this case
the slab's density is a strong function of position and the estimate of the
mass-weighted mean velocity gradient is a bit arbitrary.  Here I assumed that
the velocity gradient was constant.  A similar change takes place in
our approximation of $T_{ij}$ so that Eq. (\ref{eq:stress}) becomes
\begin{equation}
T_{ij}\approx {J_i J_j\over\Sigma}+{W^2\over 2L_s^2\Sigma}\hat n_i\hat
n_j
\label{eq:stress1}
\end{equation}

We will find it useful to rewrite Eqs. (\ref{eq:sld1}) through
(\ref{eq:stress1})
in the small bending angle limit with $D\rightarrow \infty$ and ignoring
terms describing {\it only} the secular evolution of the slab.  We find
\begin{equation}
L_s={c_s^2\Sigma\over\rho_E V_E^2},
\label{eq:zap1}
\end{equation}
\begin{equation}
\dot\Sigma=-\partial_xJ_x,
\label{eq:zap2}
\end{equation}
\begin{equation}
\dot J_x=-\partial_x\left(c_s^2\Sigma+{J_x^2\over\Sigma}+
{W^2\over 2L_s^2\Sigma}\right),
\label{eq:zap3}
\end{equation}
\begin{equation}
\dot V_z=-{c_s^2\over L_s}-{J_x\over\Sigma}\partial_x V_z-{1\over\Sigma}
\partial_x\left({W^2\partial_x (z_m+L_s)\over 2L_s^2\Sigma}\right),
\label{eq:zap4}
\end{equation}
\begin{equation}
\dot z_m=V_z-{J_x\over\Sigma}\partial_x z_m-{1\over\Sigma}\partial_x W,
\label{eq:zap5}
\end{equation}
\begin{equation}
\dot W=-W\partial_x\left({J_x\over\Sigma}\right)-W\partial_xz_m\partial_xV_z
-W\partial_x\left({J_x\over\Sigma}\right)-\Sigma c_s^2\partial_x z_m
-C_d \left|{W\over 2L_s^2\Sigma}\right|W,
\label{eq:zap6}
\end{equation}
where
\begin{equation}
V_z\equiv {J_z\over\Sigma}.
\label{eq:zap7}
\end{equation}
The critical difference between Eqs. (\ref{eq:zap1}) through (\ref{eq:zap7})
and Eqs. (\ref{eq:br1}) through (\ref{eq:wnu}) is that in this case $V_z$
is nonzero.  To be more precise, $\dot V_z$ is nonzero. For the most part
this has little effect.  In most places only the $\hat x$ derivative of
$V_z$  appears in our equations.  However, the first term on the right
hand side of Eq. (\ref{eq:zap4}) is not exclusively zeroth order.  Since
$L_s\propto\Sigma$, perturbations in the density lead directly to local
variations in $\dot V_z$ and therefore to ripples in the slab.  In this
case the breathing modes are linearly coupled to the rippling modes.
In physical terms, parts of the slab with more (less) column density undergo
less (more) deceleration relative to the slab average.  We will see
that this undercuts the NTSI, but leads instead to a family of
growing oscillatory modes.

We begin with a purely linear treatment.  To first order Eqs. (\ref{eq:zap1})
through (\ref{eq:zap7}) become
\begin{equation}
\dot\delta\Sigma=-\partial_xJ_x,
\label{eq:zapp1}
\end{equation}
\begin{equation}
\dot J_x=c_s^2\partial_x\delta\Sigma,
\label{eq:zapp2}
\end{equation}
\begin{equation}
\Delta \dot V_z=-{\delta\Sigma c_s^2\over \Sigma L_s},
\label{eq:zapp3}
\end{equation}
\begin{equation}
\Delta \dot z_m=\Delta V_z-{1\over\Sigma}\partial_x W,
\label{eq:zapp4}
\end{equation}
and
\begin{equation}
\dot W=-\Sigma c_s^2\partial_x z_m,
\label{eq:zapp5}
\end{equation}
where $\Delta V_z$ and $\Delta z_m$ are the perturbations to $V_z$ and
$z_m$.  These equations can be combined as
\begin{equation}
(\partial_t^2-c_s^2\partial_x^2)^2\Delta z_m=0.
\label{eq:lzap1}
\end{equation}
In terms of frequencies this implies that
\begin{equation}
(\omega^2-c_s^2k^2)^2=0.
\end{equation}
This dispersion relation is just
what emerges from a detailed analytical treatment of the
modes of a decelerating slab (\cite {vr89}) in the appropriate limit
(cf. appendix C).  However, it is somewhat deceptive.  The degeneracy
between the frequencies of the bending and breathing modes does
not emerge from two uncoupled sets of linear equations, but as a degeneracy
in a single coupled eigenvalue equation.  As is usual in such cases
there is a second solution proportional to $t e^{i\omega t}$.
Physically the two families of solutions correspond to pure
bending modes with $\delta\Sigma=0$ and breathing modes
which drive the bending modes resonantly.  Whether or not
this linear secular overstability actually appears depends
on the strength of linear damping terms with time scales
of order $t$.  We have ignored such terms here and their
effect will depend on the specific geometry considered.
In any case, this overstability is extremely weak and therefore of
limited interest.  However, we will see that generalizing this
instability to include nonlinear effects results in a more
robust instability.

The growing solutions to Eq. (\ref{eq:lzap1}) have the scaling
properties
\begin{equation}
\tau^{-1}\sim c_sk,
\end{equation}
\begin{equation}
\delta\Sigma\sim \Sigma k^2 L_s z_m,
\end{equation}
\begin{equation}
W\sim \Sigma c_s z_m,
\end{equation}
\begin{equation}
V_z\sim c_sk z_m,
\end{equation}
and
\begin{equation}
J_x\sim \Sigma c_sk^2 L_s z_m.
\end{equation}
If we use these scaling relations to evaluate the relative importance
of the nonlinear terms in Eqs. (\ref{eq:zap1}) through (\ref{eq:zap7})
we see that the last term in Eq. (\ref{eq:zap3}) is the dominant
nonlinear term in the linear regime.  This result isn't particularly
surprising.  The advective terms, i.e. those of the form $J_xk/\Sigma$,
are generally unimportant unless the bulk flows are of order $c_s$.
That leaves the last term in Eq. (\ref{eq:zap3}) as the only way
for the bending modes to provide feedback for the breathing modes.
Since the latter already have a linear effect on the former, the
existence of this nonlinear term raises the prospect of a closed
cycle with positive feedback.  Retaining only this nonlinear term,
we can modify Eq. (\ref{eq:lzap1}) to include quasilinear effects.
We obtain
\begin{equation}
(\partial_t^2-c_s^2\partial_x^2)^2\partial_t W={-c_s^4\over 2\Sigma L_s^3}
\partial_x^3 W^2.
\label{eq:qzap1}
\end{equation}
We note that the nonlinear feedback between the bending and breathing modes
is properly accounted for only in a fifth order equation, which suggests that
it will give rise to growing oscillations.  We will show explicitly
that this is the case.

Bounded, periodic solutions to Eq. (\ref{eq:qzap1}) are simple to find.
If we define
\begin{equation}
X\equiv {W\over\Sigma L_sc_s},
\end{equation}
and
\begin{equation}
y\equiv {1\over L_s}(x-\alpha c_s t),
\end{equation}
then if $X$ is a periodic function of $y$ Eq. (\ref{eq:qzap1}) can
be reduced to
\begin{equation}
\alpha(\alpha^2-1)^2X''={1\over 2}(X^2-<X^2>).
\label{eq:stable}
\end{equation}
We see that for a given basic periodicity in $y$ the value of
$\alpha(\alpha^2-1)^2$ scales with the amplitude of the oscillations.
This allows us to scale any specific solution to the desired
amplitude.  For example, if we assume that $X$ is periodic with
a dimensionless wavenumber $k$ then we can expand it in the form
\begin{equation}
{1\over 2}W=\cos(ky)+a_2\cos(2ky)+a_3\cos(3ky)+\ldots.
\end{equation}
We can find an approximate solution by truncating the series after $a_4$.
We obtain
\begin{equation}
{1\over 2}W\approx
W_0(\cos(ky)\mp0.372\cos(2ky)+0.107\cos(3ky)\mp0.027\cos(4ky)),
\end{equation}
and
\begin{equation}
\alpha(\alpha^2-1)^2k^2\approx \pm 0.415 W_0.
\end{equation}
This suffices to demonstrate the existence of well-behaved periodic solutions
to
Eq. (\ref{eq:qzap1}) for arbitrary $<W^2>$.  This method produces solutions
for any amplitude of $W$ for which Eq. (\ref{eq:qzap1}) is a good
approximation,
but for $z$ greater than $\sim k^2L_s^3$ $\alpha$ is no longer close to one,
i.e. the waves propagate at speeds significantly greater than $c_s$.

Unfortunately, we are actually interested in solutions that grow with time.
We expect these solutions to appear as slightly aperiodic waves.  This
expectation is based both on the form of the linear growing solution,
and also on the fact that a static ripple can be shown, after some
algebra, to produce a nonlinear effect which tends to push the ripple
back towards the shock midplane.  If we assume that $W$ has the form
$W=W(t/T;x\pm c_st)$, where $T$ is some large time depending on the
average amplitude of $W^2$ and $W$ is periodic in the second variable
with a wavelength of $2\pi/k$,  then Eq. (\ref{eq:qzap1}) implies that
\begin{equation}
(c_sk)^3 T^{-2}\sim {c_s^4 k^3 W\over \Sigma L_s^3},
\end{equation}
or
\begin{equation}
T^{-2}\sim {z_m\over L_s} \left({c_s\over L_s}\right)^2.
\label{eq:grow1}
\end{equation}
This argument doesn't prove that there are solutions that grow at
this rate, only that this likely to be rate of growth if growing
solutions exist.  Unfortunately, no simple analytic form for
$W$ corresponding to the growing solutions could be found.  The
obvious candidate, $W=(1-t/T)^{-2}X(x-c_s\alpha(t) t)$, yields
solutions to Eq. (\ref{eq:qzap1}) only for $T=\infty$ and $\alpha$
constant, and these are just the stable solutions described above.

A full integration of the relevant equations is beyond the scope of
this paper.  Instead, we will simply address the question of whether
or not the nonlinear decoupling included in Eq. (\ref{eq:qzap1}) has
a destabilizing effect on the linear modes.  We have numerically
integrated Eq. (\ref{eq:qzap1}) using dimensionless units for
which $L_s=c_s=1$ and $W$ is expressed in units of $\Sigma L_s c_s$.
Both $L_s$ and $\Sigma$ are taken to be constant in time.  This
is wrong, but is an acceptable approximation if the equations show
secular evolution at rates much faster than $t^{-1}$.  The results
are shown in Fig. 2.  The value of $W$ plotted is the spatially integrated
value of $2\langle W^2\rangle^{1/2}$.  We used an explicit first-order
difference
scheme with a constant time step of $0.01$.  Numerical instabilities
on the grid spacing scale, equal to $0.5$, were suppressed through
a fourth order smoothing routine.  The initial conditions were
set by assuming that initially $W=t\cos[k(x-c_st)]$.  The grid
was periodic with a wavelength of $100$.  The three curves correspond
to initial wavelengths of $100$ (solid line), $50$ (dotted line), and
$25$ (dashed line).  The initial evolution is not shown here.  It
simply reproduced the linear solution with the amplitude growing as $t$.

There are three points worth noting in Fig. 2.
First, the code reproduced the expected linear growth of the mode
amplitudes at early times.  Second, nonlinear effects are indeed strongly
destabilizing.  The linear solution fails, in the sense that the rate
of increase in the logarithm of $\langle W^2\rangle^{1/2}$ is at least twice
$t^{-1}$ as soon as the wave amplitude exceeds $\sim (10/t)^2$, a result
which does not depend on $k$.  Third, the subsequent evolution of the
wave amplitude is only weakly dependent on $k$, although the longer
wavelength modes do tend to lag behind somewhat during the final stages
of the nonlinear instability.  The first result is the most important
one.  It shows that although the NTSI does not operate in a cold,
decelerating, undriven slab, there is a different nonlinear instability
present.  We will refer to this as the Nonlinear Deceleration Instability
(NDI).  The second result is in agreement with Eq. (\ref{eq:grow1}), which
predicts that in the quasilinear regime the growth rate does not depend
on wavelength.  The actual growth of the mode amplitudes is faster than
the prediction of Eq. (\ref{eq:grow1}) by a logarithmic factor.  The onset
of the quasilinear regime, in terms of the necessary $z_m$ can be estimated
from our results as
\begin{equation}
z_m> 10^2 L_s\left({L_s\over c_s t}\right)\sim 10^2 L_s/D.
\end{equation}
It is sobering to remember that this coefficient of $10^2$ is a `factor
of order unity'.

The third result, that the growth of the wave amplitude in the strongly
nonlinear regime is only weakly dependent on $k$, is a little surprising.
We expect that the strongly nonlinear regime is defined by the point where
$c_skT<1$ or $W>\Sigma c_sL_s^3k^2$.  Beyond this point we expect a
growth rate of roughly $c_sk (W/W_{crit})^{1/5}$.  Clearly the larger
$k$ modes ought to have faster growth rates.  That this is not true,
or rather is realized only to a limited extent, results from
the fact that the NDI grows through the generation of strong harmonics.
In fact, by the time the strongly nonlinear regime is entered each
integration shows more power associated with the first few
harmonics  than with the original wavelength.  This leads to
a general enhancement of the growth rate of the long wavelength
modes, as well as the appearance of strong substructure.

These results are dependent on ignoring all the other nonlinear terms
in our equations.  Is this justified?  Although a detailed integration
is necessary to eliminate (or confirm) the presence of significant
effects arising from the neglected terms, simple scaling estimates of
the size of these terms indicates that only the viscous damping term
is likely to be important.  This term has a significant effect when
\begin{equation}
c_sk\left({W\over \Sigma c_sL_s^3k^2}\right)^{1/5}\sim
{|W|\over \Sigma 2L_s^2}.
\end{equation}
This condition is satisfied for
\begin{equation}
W\sim \Sigma L_s c_s(kL_s)^{3/4}.
\end{equation}
At large $W$ we can replace Eq. ({eq:qzap1}) with
\begin{equation}
\partial_t^4 z_m\approx {c_s^4\over L_s}\partial_x^2(|\partial_x z_m|).
\end{equation}
This in turn implies a modified nonlinear growth rate of
\begin{equation}
T^{-1}\sim c_s k (kL_s)^{1/4}.
\end{equation}

Apparently the asymmetric decelerating slab is subject to a previously
unknown instability, which results in growing oscillations with a
maximum growth rate of $\sim c_sk (kL_s)^{1/4}$.  This instability,
here referred to as the NDI, has an unknown threshold density contrast,
and will be particularly sensitive to numerical viscosity (like other
growing oscillatory modes).

\section{Gravitational Instabilities}

Although the bending modes of shock bounded slabs are the ones that give
rise to the NTSI, the breathing modes are not purely of academic interest.
These modes are the ones most closely connected with gravitational
instabilities.  In particular, if we consider a symmetric shock bounded
slab, whose self-gravity has a small effect on the slab structure in the
$\hat z$ direction, then gravitational clustering will create fluctuations
in $\Sigma$ and $L$, but not in the average density within the slab.
In other words, in this limit gravitational modes are simply breathing modes
with small $kL$ and a small additional force in the $\hat x$ direction.  This
force is just $2\pi G\partial_x\Sigma/k$, which changes Eq. (\ref{eq:jx2})
into
\begin{equation}
\partial_t J_x=\rho_Ec_s^2(\partial_x z_2-\partial_xz_1)-\partial_x T_{xx}
-c_s^2\partial_x\Sigma+2\pi G\Sigma\partial_x\Sigma/k.
\end{equation}
In the linear limit, with $D$ large, this becomes
\begin{equation}
\dot J_x =2\pi G\Sigma\partial_x\Sigma/k-c_s^2\partial_x\Sigma.
\end{equation}
This can be combined with Eq. (\ref{eq:lbr1}) to yield
\begin{equation}
\ddot\delta\Sigma+[c_s^2k^2-2\pi kG\Sigma]\delta\Sigma=0.
\label{eq:gdisp}
\end{equation}
This dispersion relation, including the pressure term $c_s^2k^2$, has
been previously proposed for a slab (\cite{v83,vo88}).  It can be
derived (approximately) from energy arguments (\cite{oc81}).
The derivation offered here differs from previous work by virtue
of including a detailed linear analysis of breathing modes without
vertical averaging (contained in appendix B) showing that the
pressure term is not an artifact of the thin sheet approximation.

The most interesting thing about this result it conflicts with
the dispersion relation quoted by a number of authors
(e.g. \cite{ee78,lp93}).  This disagreement follows from the
fact that they substitute external pressure
for shock boundary conditions.
Note, for example, that if we had taken thermal pressure  boundary
conditions in Eq. (\ref{eq:jx2}) we would have replaced $\rho_E$
in the first term with $\rho_{slab}$ and arrived at
\begin{equation}
\dot J_x \approx
\rho_{slab}c_s^2(\partial_xz_2-\partial_xz_1)-c_s^2\partial_x\Sigma\approx 0.
\end{equation}
In other words, pressure boundary conditions suppress transverse restoring
forces
(see also \cite{lp93}).  Shock boundary conditions do not.
The neglect of the correct boundary conditions
has been usually justified by citing linear (\cite{we82})
and nonlinear (\cite{hun86}) numerical work purporting to show that
pressure boundary conditions yield acceptable accuracy when compared
to shock confined slabs.   In the former case, Welter (1982)
noticed that the instability peaked at a wavenumber $\sim \pi G\Sigma/c_s^2$,
which is the value predicted from Eq. (\ref{eq:gdisp}) and lies
about a factor of 2 below the value predicted in Elmegreen \& Elmegreen (1978).
(In his notation, the predicted value for $kH_0$ from Eq. (\ref{eq:gdisp})
would be $10A$.  He calculated
the growth of gravitational instabilities for $A$ evolving from $0.45$
to $0.64$.  The growth rates peaked between $kH_0=5$ and $6$.)
The disagreement between different boundary conditions was as small
as a factor of two because the simulation conditions were chosen
so that the gravitational scale height was not very much larger than
the slab thickness.  Nevertheless, it is quite clear that this
calculation does not support the use of pressure boundary conditions.
The numerical study of gravitational instability in colliding flows
(\cite{hun86}) also failed to distinguish between the competing
models correctly and actually reported results close to the predictions
of Eq. (\ref{eq:gdisp}).  In any case, the significance of their result is
harder to assess since the NTSI appeared at the onset of the calculation.

One last possibility is that Eq. (\ref{eq:gdisp}) is technically correct,
but that nonlinear effects minimize the effects of the large postshock
velocities and introduce serious errors in Eq. (\ref{eq:gdisp}), possibly
leading to a situation where the dispersion relation of Elmegreen \& Elmegreen
(1978)
is approximately correct (Lubow \& Pringle 1993b).  Obviously, vertical mixing
within the slab cannot achieve this end.  Conservation of momentum
implies that the total mass flux will be unchanged by this, and in any
case such mixing has already been shown to have no effect.  The only
nonlinear process which could seriously alter Eq. (\ref{eq:gdisp})
is the cancellation of transverse momentum through collisions between
thin, rapid postshock flows.  We have already shown that shearing
instabilities should cause vertical mixing before this process can
become important.  However, it may also be useful to examine the
consequences of this argument, assuming that vertical mixing is
somehow suppressed.  The condition that transverse mixing occurs in less
than a gravitational collapse time scale is
\begin{equation}
kV_s k(\delta L)>\tau_g^{-1},
\end{equation}
or since $\delta\Sigma/\Sigma\approx \delta L/L\equiv \delta$
\begin{equation}
\delta> {1\over (kL)^2} {c_s^2\over V_s^2} {t\over\tau_g}.
\end{equation}
This is trivial to satisfy for a marginally unstable mode ($t\sim \tau_g$).
On the other hand, even if we assume that vertical mixing is suppressed,
gravitational collapse cannot proceed if the mass flux away from a
mass concentration exceeds the mass infall.  The condition that the
mass flux contained in the thin postshock layer is negligible is
\begin{equation}
V_0 {L\over kV_0 t}<{\delta\over k\tau_g} L,
\end{equation}
or
\begin{equation}
\delta>{\tau_g\over t}.
\end{equation}
For a marginally unstable mode this implies that the transverse
pressure forces implied by the use of shock boundary conditions
cannot be ignored unless all vertical mixing is suppressed {\it and} the
perturbations are already nonlinear. Evidently it is difficult
to imagine circumstances where using pressure boundary conditions
is an adequate approximation for a slab bounded by shocks.

\section{Discussion}

We have shown that slabs bounded by one or more shocks with a
high compression ratio are subject to a variety of rapidly growing
instabilities, even in the absence of linear instabilities.
For a stationary slab the relevant mechanism is the one referred to
here as the NTSI.  This instability is due to the tendency for the
postshock gas flow to transport momentum directed away from the shock
towards trailing regions of the shock.
Consequently, such slabs can be expected to
spontaneously produce substructure on scales of the slab
thickness and with characteristic velocities which are
at least of the same order as the sound speed.  The
vorticity induced by this effect is of order $c_s/L$ or
${\cal M}/t$.  A high compression ratio is necessary for this
instability for two reasons.  First, this makes the threshold
for the NTSI easily accessible in the sense that the
required initial displacement becomes small compared to the
overall dimensions of the system.  Second, this instability
requires a range of transverse wavenumbers $k$ such that
$kL$ is small and $c_skt$ is large.  More detailed
predictions for the end state of this class of instabilities
will require further numerical simulations.  The analytic model used
in this paper is only intended to follow its early growth.

A competing instability (actually a growing oscillation, or
overstability) appears when the slab is decelerating.  Here
we refer to this process as the Deceleration Overstability.
This case is somewhat more complicated, but is
basically driven by two effects. The first is the tendency of
high column density regions to move ahead of the rest of the slab.
The second is the fact that oblique sections of the shock
will generate large shears within the slab.  This shear will
act as a pressure, lowering the local column density.  The
feedback between the two effects couples the bending and
breathing modes, and produces a set of growing oscillations.
For the NDI, unlike the NTSI, there is no well defined threshold
for instability.  Instead, we go from a purely linear secular
instability to an increasingly violent nonlinear feedback.  For
typical displacements of order $k^2L_s^3$ the growth rate
surpasses $kc_s$.  At slightly higher amplitudes the growth
rate stabilizes, due to internal turbulent damping, at
approximately $c_sk (kL_s)^{-1/4}$.
Saturation sets in when the small bending limit is violated,
leading to dimensional estimates of postshock vorticity and bulk
flow similar to what we expect for the NTSI.

Our work indicates that gravitational
stability analyses of shock-bounded slabs are reliable
only to the extent that they assume the presence of a local
turbulent velocity dispersion comparable to $c_s$.  This
conclusion should have only a limited impact since a correct
derivation of the linear modes of gravitational instability
implies a similar term in the dispersion relation.

Everywhere in this paper we have assumed that the global
geometry and evolution of the shock can be neglected.  Dgani,
Walder, \& Nussbaumer (1993), and Dgani (1993) have
proposed an explanation for the results of Stevens et al.
(1992) which is based on just such effects.  They point out
that when winds from neighboring stars collide, the shocked
slab tends to flow away from the axis of symmetry, accelerated
outward by the oblique accretion of the winds.  Their analysis,
and a subsequent numerical simulation (\cite{ds93}) assumed
an infinite compression ratio so that the maximum wavelength
for the NTSI, $c_s t\rightarrow 0$.  Consequently there is
essentially no overlap between our results.  Their instability
arises from the fact that density fluctuations in the sheet
lead to fluctuations in the transverse acceleration.  It
can be interpreted as a long wavelength breathing mode of the sheet.
The resulting modes are growing oscillations with growth rates
of $\sim (kr)^{1/2} V_E/R_s$, where $r$ is the transverse
distance to the stagnation point and $R_s$ is the distance from
one of the stars to the slab.  Their treatment breaks down
when pressure effects are important, i.e. when $c_sk$ is
comparable to this growth rate.  This gives a maximum $k$
of $\sim r/(LR_s)$.  These modes are dynamically important
when they grow faster than they are advected by the transverse
expansion of the sheet, i.e. for $r>(LR_s)^{1/2}$ where
the transverse expansion rate is $\sim V_E (r/R_s)$.  They
are always traveling modes, i.e. the growth time is
never shorter than the time it takes for the gas to move
$k^{-1}$.  In general we see that these modes grow more
slowly than the NTSI, and involve longer wavelengths, unless $r$
is of order $R_s$ or more.

Although these modes undoubtedly play a role in the dynamics
of colliding winds, they cannot be identified with the
instability observed by Stevens et al. (1992) and Hunter
et al. (1986) for four reasons. First, Hunter et al.
simulated flows purely perpendicular to the cold slab.
There were no zeroth order transverse accelerations in
that case.  Second, Stevens et al. reported the
appearance of modes that grew faster than they were advected.
Third, both groups reported a rippling instability, i.e.
bending modes.  Fourth, Stevens et al. described the
growth of the instability as being centered around the
line connecting the two wind sources (the "stars'').

It is useful to contrast the instabilities
presented here with previous work on the linear overstability
of a decelerating radiative blast wave partially driven by a low density,
high pressure core (\cite{v83,vr89} and references therein).
In that work an
oscillatory instability was discovered with a maximum growth
rate close to $\sim c_s/L$ on a scale $\sim L$.  Numerical
(\cite{mn93}) and experimental (\cite{grun91}) tests of this instability
have shown that the linear theory provides a good description
up to the point where the perturbed velocities are close to the
sound speed and fractional perturbations to the column density
are of order unity (at which point the perturbations stopped
growing).  It may seem remarkable that the linear theory
could be an adequate guide in that case, whereas our treatment
here shows that smaller ripples in the shock front are
sufficient to set off nonlinear effects.  This
agreement is actually complete than it appears.
The linear treatment of this instability implies
the existence of a series of thin layers of opposing flows behind
the shock front.  For the reasons already discussed in this paper,
this should be regarded as strictly applicable only to infinitesimal
ripples.  Once the oscillations are large small scale instabilities will
blend these motions, leaving only a relatively large scale vorticity.
However, the overall growth rate is unaffected since the narrow
layers of vorticity are, in any case, non-propagating modes whose
effect on the dispersion relation is based purely on their interaction
with the compressional modes at the shock front.
Aside from this detail, the fact that the nonlinear effects
discussed here have very little effect on the linear
instability can be understood once we remember that
the growth rate of these nonlinear instabilities is (at most)
only slightly more than $c_sk$. This is typically much less than
the oscillation frequency of the overstability, which
has a characteristic time scale of roughly $c_s(k/L_s)^{1/2}$,
unless the shock front is grossly distorted.  In other words, when
the overstability is present the
geometry of the shock surface, and the slab behind it,
oscillates too fast for the nonlinear effects described in
this paper to have a dominant effect.  We note that the saturated state
of the linear overstability involves only modest changes in the
overall structure of the postshock region (aside from the
presence of turbulent velocity comparable to the sound
speed).  On the other hand, the saturated state of the
NTSI and NDI are not known, and may involve
gross distortions of the slab position.  It is even possible
that the linear overstability actually tends to {\it stabilize}
the radiative blast wave, in the sense that it may produce smaller
distortions than those which would otherwise be induced by
nonlinear effects.

How general is the presence of nonlinear instabilities?
For a slab bounded on one side
by a shock, and on the other by a diffuse gas with a large pressure,
we have found stability when the slab is stationary.  However, this
is probably a very special case.  When the thermal pressure is
stronger the slab is Rayleigh-Taylor unstable.  When the thermal
pressure is smaller the slab is subject to growing oscillations.
When the thermal pressure is small our work in section 3B shows
that the NDI will appear.  Given this, the appearance of
strong postshock substructures seems inevitable.  When a slab
is bound by two shocks we see that for the case where the ram pressures are
equal, and the case where one is much larger, nonlinear instabilities
appear.  Following the methods used in this paper it is straightforward
to show that the intermediate cases are similarly unstable.

One of the more speculative applications of this work would be
to the formation of galaxies within `pancakes' of shocked gas
(\cite {z170,z270}).  In this model scalar cosmological
perturbations lead
to the formation of caustics in the gas flow.  Initially these
are quite similar to the shock confined slabs discussed above.
However, after some time the cooling of the postshock gas becomes
delayed and the cold slab gradually builds up surrounding layers
of hot gas.  Moreover as the column density of the slab increases
gravitational instabilities will become important.  The gravity
due to the slab will tend to suppress the instability by acting
to pull everything towards the midplane.  A crude estimate of this
effect can be obtained from comparing the vertical gravity of
the slab, $2\pi G\Sigma$, with the upward acceleration of a bend
in the slab, $\eta (c_sk)^2 k\eta$.  Assuming that the NTSI
becomes ineffective in the limit of strong bending we have
$(k\eta)^2<1$ so that the condition that the gravity of the
slab is negligible is
\begin{equation}
2\pi G\Sigma< (k\eta)^2 c_s^2 k < c_s^2 k
\label{pancak}
\end{equation}
However, the maximum wavenumber for gravitational instability in
a thin slab bounded by shocks is just $2\pi G\Sigma/c_s^2$ (\cite{v83};
also section 4 of this paper)
so Eq. (\ref{pancak}) is equivalent to assuming that the NTSI is
important on all scales between the slab thickness and the wavelength
at which gravitational instabilities are important.  This
qualitative estimate is not sensitive to the gravitational effects of dark
matter,
since its presence will increase the suppression of the NTSI to the
same extent that it lowers the minimum wavelength of gravitational
instability. We conclude that the NTSI guarantees that the cold slab begins as
a turbulent layer with a strong local vorticity.  Among other things,
this implies that gravitational instabilities will have seeds that
are already of order unity.  The fragments generated in this
stage of its evolution will tend to have rotation axes that lie
in the in the plane of the pancake.  We note that the angular momentum
generated in this way is present at earlier times, and on smaller
scales, than the large scale shearing instability mechanism proposed by
Doroshkevich (1971, see also \cite{bi74}).  Their later evolution will
depend on the specific model of pancake formation.  Preliminary attempts
to model pancake instabilities (\cite{y91}) show that the emergent
substructure is sensitive to the presence of cooling instabilities.
However, it is important to stress that the effects of
the NTSI can only be evaluated only by numerical simulations
that, unlike Yuan et al., do not impose a mirror symmetry around
the midplane.

The fact that the NTSI and NDI are strong sources of vorticity implies
that any volume of gas subjected to strongly compressive
shocks will contain strong local eddys, a result which would
not follow from purely gravitational forces.  Examples
would include very young galaxies and star formation
regions today.  This in turn implies that given any pre-existing
magnetic field at all such regions will quickly evolve to the
point where they contain dynamically significant magnetic
fields on scales at least as large as the typical eddy scale.
This effect is clearly relevant to dynamo models for galactic
magnetic fields, although it is still uncertain whether or not
such models are physically realistic. This work implies that
galaxies will contain strong small scale fields from the first
epoch of star formation (or earlier, depending the kinds of
gas flows present during galaxy formation).  However, the major
stumbling block in such theories is the lack of realistic mechanism
for transferring small scale magnetic field energy to galactic
scales (e.g. \cite{ka92}).  In any case
in regions of active star formation this effect
will dominate over the ejection of magnetic flux from stars, a process
which only gives small field strengths when diluted over typical
interstellar distances (\cite {r87}).  The efficient generation of vorticity,
and the consequent rapid amplification of local magnetic fields,
provides a physical basis for the suggestion
(\cite {pk89,cw93}) that turbulence in star forming regions
is an important element in generating interstellar magnetic fields.
We note in passing that Klein, McKee, and Colella (1993) have
studied numerically the case of a steady shock interacting with a small, dense
cloud, and find that in this case too there is efficient generation
of vorticity.  In some sense that work is complementary to this paper,
in that it shows that in an inhomogeneous medium the growth of shock
wave instabilities from small perturbations will compete with rarer,
but more intense, events caused by exceptional external inhomogeneities.

{}From an observational point of view, the most significant implication
of this work is that strongly cooling shocks will typically have
extremely corrugated surfaces on scales comparable to, or larger
than, the cold slab thickness.  This means that as gas crosses
the shock front it will retain a large fraction of its original
speed, which will now have a significant component in the plane
of the cold slab.  This remaining momentum will be dissipated
through a series of secondary shocks in a supersonic turbulent
flow.  Observations of emission lines excited at the shock surfaces
will show a large amount of Doppler broadening, somewhat less than
the preshock velocity, and may be excited repeatedly in the secondary
shocks.  Lines excited further `downstream' in the flow will
show smaller velocity dispersions.  Line emission characteristic
of the bulk of the cold slab will reveal velocity dispersions comparable
to the local sound speed.

It is also tempting to try to relate this work to the well known
tendency of old shock waves to show filamentary substructure.
Mac Low and Norman (1993) suggested that these filaments
might be the result of viewing shock front ripples obliquely,
so that whenever the line of sight is tangent to the shock
a `filament' appears.  Their suggestion was made in the context
of the linear instability, but applies equally well here.
However, in order for long filaments to appear there has to
be a tendency for a wave in one direction to interfere
with the formation of waves pointing in other directions.
Demonstrating such an effect would require a three dimensional
analysis of the saturated dynamical state of the cold slab.

\acknowledgements

I am happy to acknowledge discussions with John Blondin
in which he goaded me to think about this problem.  Ruth
Dgani, Patrick Diamond, Daniel
Hiltgen, John Lacy, Steve Lubow, John Pringle, John Scalo,
and Paul Shapiro also contributed useful comments.

This work was supported in part by NSF grant AST-9020757.

\appendix
\section{Derivation of the Shock Boundary Conditions}

We consider a slab bounded by isothermal shocks whose position is
given by the functions $z_2(x,t)$ and $z_1(x,t)$.  The gas outside
the slab has density $\rho_E$ and is moving with a velocity
$\pm V_E\hat z$.  In the frame in which a shock is stationary
and the preshock gas has a velocity perpendicular to the
shock of $v_{\perp}$ and a velocity tangential to the shock
of $v_{T}$ the postshock gas has a density of
\begin{equation}
\rho_s=\rho_E\left({v_{\perp}\over c_s}\right)^2,
\end{equation}
a perpendicular velocity of
\begin{equation}
v_{s,\perp}={c_s^2\over v_{\perp}},
\end{equation}
and a tangential velocity which is unchanged.

Let's start by considering the shock at $z_2(x,t)$ (the other
set of conditions will follow from a trivial change in symmetry).
In this case the unit vector normal to the shock, $\hat n_\perp$,
is
\begin{equation}
\hat n_\perp\equiv {\hat z-\partial_x z_2\hat x/over
[1+(\partial_xz_2)^2]^{1/2}}.
\end{equation}
It will also be useful to define the unit vector tangential
to the shock, $\hat n_T$, which is
\begin{equation}
\hat n_T\equiv {\hat x+\partial_x z_2\hat z/over
[1+(\partial_xz_2)^2]^{1/2}}.
\end{equation}
Of course, this definition includes an arbitrary choice of
sign, which will not affect our final results.
Next we note that the velocity of the shock wave is,
by definition, purely in the direction of $\hat n_\perp$
so that it is given by
\begin{equation}
\vec v_{front}={\dot z_2\over [1+(\partial_xz_2)^2]^{1/2}}
\hat n_{\perp}.
\end{equation}
Therefore, the speed with which the gas collides with the
shock front, in the frame of the shock front, is
\begin{equation}
v_\perp={V_E+\dot z_2\over [1+(\partial_xz_2)^2]^{1/2}}.
\end{equation}
The tangential gas speed is just
\begin{equation}
v_T={V_E\partial_x z_2\over[1+(\partial_xz_2)^2]^{1/2}}.
\end{equation}

The gas properties are then transformed across the shock
front by the shock jump conditions given above.  We find that
\begin{equation}
\rho_s=\rho_E\left({V_E+\dot z_2\over c_s}\right)^2
[1+(\partial_x z_2)^2]^{-1}.
\end{equation}
The postshock gas speed, in the shock frame, is
\begin{equation}
\vec v_s={c_s^2\over V_E+\dot z_2}
[1+(\partial_x z_2)^2]^{1/2}\hat n_\perp
+{V_E\partial_x z_2\over[1+(\partial_x z_2)^2]^{1/2}}
\hat n_T.
\end{equation}
This can be transformed to the lab frame by adding
$\vec v_{front}$ to $\vec v_s$.  Using our definitions
of $\hat n_\perp$ and $\hat n_T$ we obtain the postshock
gas speed at $\vec z_2$.  It is
\begin{equation}
\vec v(z_2)=\left({\dot z_2\over 1+(\partial_xz_2)^2}-
{c_s^2\over V_E+\dot z_2}\right)(\hat z-\partial_x z_2\hat x)
- {\partial_x z_2V_E\over 1+(\partial_xz_2)^2}(\hat x+\partial_x z_2\hat z).
\end{equation}

The corresponding conditions for $\vec z_1$ can be obtained
through the substitution of $z_1$ for $z_2$ and $-V_E$
for $V_E$.

\section{Linear theory for the symmetric slab without vertical averaging}

Although vertical averaging allows us to follow the slab evolution
into the nonlinear regime, it is unnecessary for purely linear
perturbations.  Here we show the linear perturbation treatment
for the shock bounded slab as a test of our vertically averaged
results, in particular our claim that the bending modes of such a slab
are stiff.

We start by noting that the linear solutions within the slab will
contain a compressional piece, a sound wave, satisfying the
usual wave equation, and a nonpropagating vortical part.  Only
the combination of the two will satisfy the linear
boundary conditions.  From Eqs. (\ref{eq:vj2}), (\ref{eq:rj2}),
(\ref{eq:vj1}), and (\ref{eq:rj1}) we see that these are
\begin{equation}
\vec v(z_2)=\Delta\dot z_2\left(1+{c_s^2\over V_s^2}\right)\hat z
-V_E\partial_x\Delta z_2\hat x,
\end{equation}
\begin{equation}
\rho(z_2)=\left({V_s\over c_s}\right)^2\rho_E
\left(1+2{\Delta \dot z_2\over V_s}\right),
\end{equation}
\begin{equation}
\vec v(z_1)=\Delta\dot z_1\left(1+{c_s^2\over V_s^2}\right)\hat z
+V_E\partial_x\Delta z_1\hat x,
\end{equation}
and
\begin{equation}
\rho(z_1)=\left({V_s\over c_s}\right)^2\rho_E
\left(1-2{\Delta \dot z_1\over V_s}\right).
\end{equation}

Let's start with the bending modes of the slab.  In this case
$\Delta z_2=\Delta z_1\equiv\eta$.  We will assume that the
perturbation variables are all proportional to $e^{ikx}$. The
boundary conditions
for the density perturbations can only be satisfied by the
sound wave.  We have
\begin{equation}
\delta P\approx {4\dot\eta\over V_s}\left({z\over L}\right)P_0,
\end{equation}
where $P_0$ is the unperturbed pressure.  We have neglected higher
order terms in $z$ (the next will be proportional to $z^3$).  These may
be recovered once the exact evolution of the solution is known.
However, as long as $\omega^2\approx c_s^2k^2$, which turns out
to be the case, these corrections are of little interest.
Remembering that $L\propto t$
and assuming that the oscillation frequency is much greater than $t^{-1}$
we have
\begin{equation}
v_{c,z}\approx -{4c_s^2 \eta\over V_sL}\left(1+{1\over i\omega t}\right),
\end{equation}
and
\begin{equation}
v_{c,x}\approx +{4ik \eta zc_s^2\over V_sL}\left(1+{1\over i\omega t}\right),
\end{equation}
where the subscript $c$ indicates that these solutions are only the compressive
part of the full velocity field.  In deriving these results we have assumed
that $\eta\propto e^{i\int^t\omega dt}$.
The condition that the remaining pieces of the velocity field are purely
vortical can be expressed as
\begin{equation}
0=ik\left[-V_Eik\eta-{2ik\eta c_s^2\over V_s}\left(1+{1\over i\omega
t}\right)\right]
+2\dot L^{-1}\partial_t\left[\dot\eta\left(1+{c_s^2\over
V_s^2}\right)+{4c_s^2\eta\over V_s L}
\left(1+{1\over i\omega t}\right)\right],
\end{equation}
where we have used the fact that the vorticity perturbations are nonpropagating
to
substitute $2\dot L^{-1}\partial_t$ for $\partial_z$.
Discarding terms of higher order than $c_s^2/V_s^2$ we find that this implies
that the bending
modes have the form
\begin{equation}
\eta\propto {1\over t}e^{i(kx\pm \omega t)},
\end{equation}
where
\begin{equation}
\omega=c_sk\left(1-{2c_s^2\over V_s^2}\right)
\end{equation}
We note that $|\dot \eta |\gg |v_{c,z}|$, i.e. the surface fluctuations
associated
with the bending modes are not
driven by the bulk motion of the slab, but by the tendency of freshly accreted
material to focus on concavities and move away from convexities.

The breathing modes of the slab have $\Delta z_2=-\Delta z_1\equiv\eta$.
Consequently
\begin{equation}
\delta P\approx {2\dot\eta\over V_s}P_0,
\end{equation}
and
\begin{equation}
v_{c,x}\approx -{2ik \eta c_s^2\over V_s}.
\end{equation}
Although $v_{c,z}$ is not precisely zero, it is of higher order and will be
neglected in what follows.
Once again applying the boundary conditions and the incompressible nature of
the vortical perturbations we find that
\begin{equation}
0=ik\left[-V_Eik\eta+{2ik\eta c_s^2\over V_s}\right]
+2\dot L^{-1}\partial_t\left[\dot\eta\left(1+{c_s^2\over V_s^2}\right)\right].
\end{equation}
This implies that the solutions to the breathing modes have the form
\begin{equation}
\eta\propto e^{i(kx\pm \omega t)},
\end{equation}
where
\begin{equation}
\omega=c_sk\left(1-{2c_s^2\over V_s^2}\right)
\end{equation}

\section{Linear theory for asymmetric slabs}

We can derive the linear modes of asymmetric slabs by taking the
appropriate limit of the dispersion relation given in Vishniac
and Ryu (1989) for a decelerating slab bounded on one side
by a hot, diffuse medium and on the other side by a shock with
a very high density contrast.  We have
\begin{equation}
\omega^2(\omega^2-c_s^2k^2)+{k^2c_s^4\over L_s^2}\left({1+\beta^Q\over 1-
\beta^Q}Q-1\right)=0,
\label{eq:dispap}
\end{equation}
where $\beta$ is the ratio of the thermal pressure to the ram pressure,
$L_s$ is the density scale height in the slab, and
\begin{equation}
Q\equiv \left[1+4(kL_s)^2-{4L_s^2\omega^2\over c_s^2}\right]^{1/2}.
\end{equation}
Due to a typographical error
the factors of $4$ are missing from Eq. (8b) in Vishniac and Ryu (1989).
These factors subsequently reappear in their treatment so their results
are not compromised in any way.
We note that as long as $\beta\ne0$ the sign of $Q$ is irrelevant.
The density scale height, $L_s$, and $\beta$ are not completely
independent, in the sense that the column density is given
by
\begin{equation}
\Sigma=\rho_s L_s(1-\beta),
\label{eq:constraint}
\end{equation}
where $\rho_s$ is the density just behind the shock front.

A stationary slab bounded on one side by a shock, and on the other
side by thermal pressure has $\beta=1$ and $L_s\rightarrow\infty$.
{}From Eq. (\ref{eq:constraint}) this implies that in this limit
\begin{equation}
(1-\beta)L_s\rightarrow L.
\end{equation}
If $\omega^2=c_s^2k^2$ then $Q=1$ and
\begin{equation}
{k^2c_s^4\over 2L_s^2}\left({1+\beta\over1-\beta}-1\right)\rightarrow
{k^2c_s^4\over LL_s}\rightarrow 0.
\end{equation}
So in this case the second term in Eq. (\ref{eq:dispap}) vanishes.
However, for $\omega^2=c_s^2k^2$ the first term also vanishes and
the dispersion relation is satisfied.

Suppose $\omega^2\ne c_s^2k^2$?  Then
\begin{equation}
Q\rightarrow 2L_s\left(k^2-{\omega^2\over c_s^2}\right)^{1/2}.
\end{equation}
Then
\begin{equation}
\beta^Q=\exp(Q\ln\beta)\rightarrow \exp[-2H(k^2-\omega^2/c_s^2)^{1/2}],
\end{equation}
which is not equal to zero or one.  Therefore
\begin{equation}
{k^2c_s^4\over 2L_s^2}\left({1+\beta^Q\over 1-\beta^Q}Q-1\right)\rightarrow
\left({1+\beta^Q\over 1-\beta^Q}\right){k^2c_s^2\over L_s}\left(k^2-{\omega^2
\over c_s^2}\right)^{1/2}\rightarrow 0.
\end{equation}
This implies that
\begin{equation}
\omega^2(\omega^2-c_s^2k^2)=0.
\end{equation}

We conclude that the stationary asymmetric slab has linear modes with
\begin{equation}
\omega^2=0, c_s^2k^2,
\end{equation}
in agreement with the results of section 3.1.

The asymmetric slab with no pressure boundary at all is described by
Eq. (\ref{eq:dispap}) in the limit with $\beta\rightarrow 0$.
This is
\begin{equation}
\omega^2(\omega^2-c_s^2k^2)+{k^2c_s^4\over L_s^2}(Q-1)=0,
\label{eq:dispap2}
\end{equation}
with Re$(Q)>0$.  Switching $Q$ to the right hand side and squaring
both sides we find that Eq. (\ref{eq:dispap2}) reduces to
\begin{equation}
(\omega^2-c_s^2k^2)^2\left(\omega^4-{k^2c_s^4\over L^2}\right)=0.
\end{equation}
The existence of two families of solutions with $\omega^2=c_s^2k^2$ is
not a surprise.  However the third family has $\omega^2=\pm c_s^2k/L$,
suggesting the presence of a linear instability.  We can substitute
this solution back into Eq. (\ref{eq:dispap2}).  We obtain
\begin{equation}
\omega^2={c_s^2\over 2L_s^2}(Q+1).
\end{equation}
This shows that $\omega^2>0$ and there is no linear instability.  For
this family of solutions
\begin{equation}
Q=\left(1+4(kL_s)^2-4kL_s\right)^{1/2}=\pm(1-2kL_s),
\end{equation}
where the sign is chosen to obtain a positive $Q$.  We see
that $\omega^2=c_s^2k/L_s$ is a solution to the dispersion
relation only for $k>L_s/2$, i.e. in the short wavelength
limit.

\clearpage
\begin{figure}
\plotone{bent.eps}
\caption{A shock confined slab with a large bending mode.  Note that
postshock flows move vertical momentum into the extrema.}
\end{figure}
\begin{figure}
\plotone{grow.eps}
\caption{The growth of the mean wave amplitude for the asymmetric, decelerating
slab.  Initial wavelengths are $\lambda= 100, 50,$ and $25$.}
\end{figure}
\end{document}